\documentclass[11pt,a4paper]{article}
\pdfoutput=1

\usepackage{jheppub}

\usepackage{amsmath, amssymb, amsthm} 
\usepackage{mathtools}
\usepackage{thmtools}
\usepackage{bm}
\usepackage{dsfont}
\usepackage{braket}

\numberwithin{equation}{section}

\usepackage{enumitem}
\setlist[itemize]{noitemsep}
\setlist[description]{noitemsep}

\usepackage[dvipsnames]{xcolor}
\usepackage{tikz}
\usepackage{pgfplots}
\usetikzlibrary{intersections,backgrounds}
\usepgfplotslibrary{fillbetween}
\pgfplotsset{compat = newest}

\usepackage[numbers, sort&compress]{natbib}
\usepackage{hyperref}
\hypersetup{colorlinks=true, linktoc=page, linkcolor=purple, citecolor=blue}

\newcommand{\cA}{\mathcal{A}}

\newcommand{\cN}{\mathcal{N}}
\newcommand{\cO}{\mathcal{O}}

\newcommand{\cW}{\mathcal{W}}

\newcommand{\bbC}{\mathbb{C}}

\newcommand{\bbN}{\mathbb{N}}

\newcommand{\bbZ}{\mathbb{Z}}

\newcommand{\lam}{\lambda}

\renewcommand{\[}{\left[}

\DeclareMathOperator*{\Res}{Res}

\newcommand{\floor}[1]{\cramped{\left\lfloor #1 \right\rfloor}}

\newcommand{\half}{\tfrac{1}{2}}

\newcommand{\negphantom}[1]{
    \ifmmode\settowidth{\dimen0}{$#1$}
    \else\settowidth{\dimen0}{#1}
    \fi
    \hspace*{-\dimen0}}
    
\makeatletter
\newcommand{\mask}[2]{{\mathpalette\mask@{{#1}{#2}}}}
\newcommand{\mask@}[2]{\mask@@{#1}#2}
\newcommand{\mask@@}[3]{%
  \settowidth{\dimen@}{$\m@th#1#2$}%
  \makebox[\dimen@]{$\m@th#1#3$}%
}
\makeatother
    
\newcommand{\sm}{\smallskip}
\newcommand{\no}{\nonumber}

\begin{document}

\title{Generalized Veneziano and Virasoro amplitudes}

\date{\today}

\author[a]{Nicholas Geiser}
\author[a]{and Lukas W. Lindwasser}

\affiliation[a]{
    Mani L. Bhaumik Institute for Theoretical Physics\\
    Department of Physics and Astronomy\\
    University of California, Los Angeles, CA 90095, USA}

\emailAdd{ngeiser@physics.ucla.edu}
\emailAdd{lukaslindwasser@physics.ucla.edu}


\abstract{We analyze so-called generalized Veneziano and generalized Virasoro amplitudes. Under some physical assumptions, we find that their spectra must satisfy an over-determined set of non-linear recursion relations. The recursion relation for the generalized Veneziano amplitudes can be solved analytically and yields a two-parameter family which includes the Veneziano amplitude, the one-parameter family of Coon amplitudes, and a larger two-parameter family of amplitudes with an infinite tower of spins at each mass level. In the generalized Virasoro case, the only consistent solution is the string spectrum.}


\maketitle


\section{Introduction}
\label{sec:intro}

In this paper, we shall search for consistent generalizations of the Veneziano amplitude~\cite{Veneziano:1968yb} and the Virasoro amplitude~\cite{Virasoro:1969me} with zero Regge intercept. For simplicity we shall only consider the scattering of four massless bosonic states, in which case the tree-level open and closed superstring amplitudes respectively reduce to the Veneziano and Virasoro amplitudes with zero intercept. Both amplitudes may be written as infinite products with an infinite number of simple poles.

\sm

So-called \textit{generalized Veneziano amplitudes} and \textit{generalized Virasoro amplitudes} are defined by modifying these infinite products subject to some general physical constraints. The name generalized Veneziano amplitude originates in~\cite{Fairlie:1994ad}, and the Coon amplitude~\cite{Coon:1969yw} is one well-studied example. In a companion paper~\cite{Geiser:2022icl}, we detail the properties of the Veneziano, Virasoro, and Coon amplitudes, including their unitarity, high-energy behavior, low-energy expansion, and number theoretic properties.

\sm

Our present procedure is an extension and clarification of Coon's original argument~\cite{Coon:1969yw} and related work~\cite{Fairlie:1994ad}. These previous studies only considered generalized Veneziano amplitudes, but we shall also consider the generalized Virasoro case. In either case, we simply assume crossing symmetry in the Mandelstam variables and demand physical residues on an a priori unspecified sequence of poles~$\lambda_n$. In other words, we do not assume the mass spectrum of the theory. Under our assumptions, we find that the poles~$\lambda_n$ must satisfy an over-determined set of non-linear recursion relations. These recursion relations fix all the subsequent poles in terms of the first three poles and highly constrain the space of generalized Veneziano and generalized Virasoro amplitudes.

\sm

In the generalized Veneziano case, the recursion relations can be solved analytically. The solutions correspond to the Veneziano amplitude, the one-parameter family of Coon amplitudes, and a larger two-parameter family of amplitudes with an infinite tower of spins at each mass level. This two-parameter family of solutions has been previously identified~\cite{Coon:1969yw, Fairlie:1994ad} but never systematically studied. Only the one-parameter sub-family of Coon amplitudes has been studied in detail~\cite{Figueroa:2022onw, Geiser:2022icl}. In this paper, we shall systematically analyze the entire two-parameter space of generalized Veneziano amplitudes. We also begin an initial study of the unitarity properties of this space

\sm

In the generalized Virasoro case, we numerically demonstrate that the only consistent solution to the aforementioned recursion relations is the string spectrum. That is, we do not find any consistent generalized Virasoro amplitudes beyond the Virasoro amplitude itself. We reached a similar, though less general, conclusion in~\cite{Geiser:2022icl} by failing to construct a generalization of the Virasoro amplitude with the same poles as the Coon amplitude (a so-called Virasoro-Coon amplitude).

\sm

The authors of~\cite{Cheung:2022mkw} approach this same problem under a different set of assumptions and reach many of the same conclusions that we reach here, such as the uniqueness of the Virasoro amplitude. Our work is complementary.

\sm

Our approach is part of the modern S-matrix bootstrap program~\cite{Correia:2020xtr}, a revival of an old approach~\cite{Eden:1966dnq} which attempts to construct general amplitudes which satisfy various physical properties without relying on an underlying dynamical theory.

\subsection{Conventions}

In this paper, we shall only consider crossing-symmetric tree-level scattering amplitudes for four massless external particles in weakly-coupled theories in~${d \geq 3}$ spacetime dimensions. We use units in which the lowest massive state of any particular theory has mass~${m^2 = 1}$. In open (closed) string theory, this choice corresponds to~${\alpha' = 1}$ (${\alpha' = 4}$).

\subsubsection{Kinematics}

We shall primarily consider amplitudes stripped of their dependence on the polarizations or colors of the scattered states, leaving functions~$\cA(s_{ij})$ which depend only on the Mandelstam variables~${s_{ij} = -(p_i+p_j)^2}$,
\begin{align}
    s
    &=
    s_{12}
    =
    s_{34}
    =
    \phantom{-} 4 E^2
    \phantom{(1-\cos\theta)}
    \geq 0
\no \\
    t
    &=
    s_{14}
    =
    s_{23}
    =
    -2 E^2 (1-\cos\theta)
    \leq 0
\no \\
    u
    &=
    s_{13}
    =
    s_{24}
    = -2 E^2 (1+\cos\theta)
    \leq 0
\end{align}
which satisfy the mass-shell relation~${s+t+u=0}$. Here~$E$ and~$\theta$ are the center-of-mass energy and scattering angle, respectively. The inequalities refer to the physical scattering regime with real~$s_{ij}$.

\subsubsection{Crossing}

Crossing symmetry refers to permutation symmetry in $(s,t)$ or $(s,t,u)$. The Veneziano, Virasoro, and Coon amplitudes were discovered under the assumption of crossing symmetry, and we are searching for their generalizations.

\sm

Since $s$-channel and $t$-channel Feynman diagrams correspond to the same cyclic ordering, color-ordered amplitudes (e.g. gluon amplitudes) will have only $s$-channel poles and $t$-channel poles and shall be denoted by~$\cA(s,t)$ to emphasize that they are analytic functions of two complex variables. For these amplitudes, crossing symmetry is the requirement that $\cA(s,t) = \cA(t,s)$.

\sm

Amplitudes with poles in all three channels (e.g. graviton amplitudes) shall be denoted by~$\cA(s,t,u)$. We shall regard these amplitudes as analytic functions of three complex variables restricted to the algebraic variety defined by~$s+t+u=0$. For these amplitudes, crossing symmetry is the requirement that $\cA(s,t,u) = \cA(\sigma(s),\sigma(t),\sigma(u))$ for any permutation $\sigma$ of $(s,t,u)$.

\subsubsection{Analytic structure}
\label{sec:analytic}

The amplitude~$\cA(s_{ij})$ is an analytic function of the complexified $s_{ij}$ with simple poles and branch cuts dictated by unitarity. At high-energy, we demand that ${\cA(s_{ij}) \to 0}$ vanishes as~${|s| \to \infty}$ with physical~$t$, in analogy with the high-energy behavior of the Veneziano, Virasoro, and Coon amplitudes~\cite{Geiser:2022icl}. 

\sm

Tree-level amplitudes have simple poles at~${s_{ij} = m_n^2}$ for each state~$n$ which couples to the external states through the $s_{ij}$-channel. It is often assumed that physical tree-level amplitudes are meromorphic, i.e.\ that~$\cA(s_{ij})$ is analytic outside its simple poles with no branch cuts or other singularities. However, the Coon amplitude with~${q<1}$ provides a counterexample of a seemingly healthy non-meromorphic tree-level amplitude~\cite{Figueroa:2022onw, Geiser:2022icl}.

\sm

In a physical four-point amplitude, the $t$-channel and $u$-channel poles should cancel on each $s$-channel pole (and vice versa). Typically, the residue of each $s$-channel pole is then a polynomial in~$t$ (after using the mass-shell relation to eliminate any $u$-dependence). The highest power of $t$ in this residue corresponds to the highest-spin state exchanged on that pole. Non-polynomial residues can in principle result and may be Taylor expanded, corresponding to the exchange of an infinite tower of spinning states. In any case, the residues of these poles can be written as a sum of Gegenbauer polynomials and the amplitude may be written as follows (under the assumption that $\cA(s_{ij})$ vanishes at high-energy~\cite{Caron-Huot:2016icg}),
\begin{align}
\label{eq:partialwave}
    \cA(s_{ij})
    =
    \sum_{n}
    \frac{ 1 }{ s-m_n^2 }
    \sum_{j}
    c_{n,j} \, 
    C_j^{(\frac{d-3}{2})}(\cos\theta)
\end{align}
In a unitary theory, the partial wave coefficients $c_{n,j}> 0$ will be positive.

\sm

We shall restrict our discussion to amplitudes with an infinite number of simple poles (\`a~la Veneziano, Virasoro, and Coon) because the assumptions of crossing symmetry, physical residues, and that~$\cA(s_{ij})$ vanishes at high-energy imply that there must be an infinite number of poles in each channel~\cite{Camanho:2014apa, Caron-Huot:2016icg}. The argument may be summarized as follows. A crossing-symmetric tree-level amplitude ${\cA(s,t)=\cA(t,s)}$ may be expanded on either its $s$\nobreakdash-channel or $t$\nobreakdash-channel poles, leading to the following equality,
\begin{align}
\label{eq:crossing}
    \sum_{n}
    \frac{ f_n(t) }{ s-m_n^2 }
    =
    \sum_{n}
    \frac{ f_n(s) }{ t-m_n^2 }
\end{align}
The functions $f_n(z)$ must be finite at each ${z = m_n^2}$ because the $t$\nobreakdash-channel poles should cancel on each $s$\nobreakdash-channel pole. However, the left-hand side of~\eqref{eq:crossing} can then only produce the $t$\nobreakdash-channel poles which appear on the right-hand side if the sum over $n$ is infinite.

\subsubsection{Accumulations points}

In this paper, we shall encounter two distinct notions of accumulation point spectra:
\begin{itemize}
    \item infinite tower of masses $m_n^2 < \lambda_\infty$ with \textit{finite spin exchange} at each mass level
    \item infinite tower of masses $m_n^2 < \lambda_\infty$ with \textit{infinite spin exchange} at each mass level
\end{itemize}
for some finite accumulation point of masses $0 <  \lambda_\infty < \infty$.

\sm

Finite spin exchange results from a polynomial residue on a given mass pole and corresponds to a finite tower of states at that mass level. The Coon amplitude with~${q<1}$ exhibits this type of accumulation point spectrum with ${\lambda_\infty = \frac{1}{1-q}}$. While there is yet no definitive physical realization of the Coon amplitude, similar accumulation point spectra have been found in a stringy setup involving open strings ending on a D-brane~\cite{Maldacena:2022ckr}. Most famously, the hydrogen atom has a spectrum of this type with energy levels ${E_n = -13.6 \text{ eV} / n^2}$ and an accumulation point at~${E_\infty = 0}$.

\sm

Infinite spin exchange results from a non-polynomial residue on a given mass pole and is generally considered unphysical. Indeed, sensible quantum field theories are typically assumed to have a finite number of particle types below any finite mass. In the case of finite spin exchange, this assumption only fails at masses ${m^2 \geq \lambda_\infty}$. In the case of infinite spin exchange, this assumption fails at all masses above the mass gap. Nevertheless, amplitudes with infinite spin exchange were recently considered in~\cite{Huang:2022mdb}. Moreover, amplitudes with this type of accumulation point were recently found to have interesting extremal properties in the context of the EFT-hedron~\cite{Caron-Huot:2020cmc, Arkani-Hamed:2020blm, Huang:2020nqy, Bern:2021ppb}.

\sm

In any case, amplitudes with either type of accumulation point spectra are not well understood and are fruitful examples for the study of general scattering amplitudes.

\subsection{Outline}

In~\autoref{sec:review}, we shall briefly review our conventions for the Veneziano, Virasoro, and Coon amplitudes. In~\autoref{sec:infprod}, we review some complex analysis and motivate our infinite product ansatz for the generalized Veneziano and generalized Virasoro amplitudes. In~\autoref{sec:Ven} and~\autoref{sec:Vir}, we respectively analyze the generalized Veneziano and generalized Virasoro amplitudes by solving an infinite set of non-linear constraints on their poles $\lam_n$. Finally, in~\autoref{sec:disc}, we discuss our results and present some questions for future research. 

\subsection*{Acknowledgements}

We are grateful to (in alphabetical order) Maor Ben-Shahar, Clifford Cheung, Eric D'Hoker, Enrico Herrmann, Callum Jones, Dimitrios Kosmopoulos, Per Kraus, Grant Remmen, Joao Silva, Oliver Schlotterer, and Terry Tomboulis for various discussions related to this work. NG is supported by a National Science Foundation (NSF) grant supplement from the Alliances for Graduate Education and the Professoriate Graduate Research Supplements (AGEP-GRS). NG was also supported in part by the NSF under Grant No. NSF PHY-1748958 and the hospitality of the Kavli Institute for Theoretical Physics. NG and LL are supported by the Mani L. Bhaumik Institute for Theoretical Physics. 


\section{Veneziano, Virasoro, and Coon amplitudes}
\label{sec:review}

The Veneziano, Virasoro, and Coon amplitudes are each tree-level four point amplitudes with an infinite sequence of simple poles and polynomial residues. A detailed review of their properties may be found in~\cite{Geiser:2022icl}. The Coon amplitude was also recently discussed in~\cite{Figueroa:2022onw, Chakravarty:2022vrp, Bhardwaj:2022lbz}. Here we shall briefly review our conventions and give each amplitude's infinite product representation.

\subsection{Veneziano}

The Veneziano amplitude~$\cA_{\text{Ven}}$ describes the scattering of four open strings and is a UV-completion of maximally supersymmetric Yang-Mills field theory. The color-stripped tree-level field theory amplitude which describes the scattering of any four massless particles in the Yang–Mills supermultiplet is,
\begin{align}
\label{eq:SYM}
    \cA_{\text{SYM}}
    &=
    P_4 \, \frac{1}{st}
\end{align}
where~$P_4 = \cO(s,t)^2$ is a kinematic pre-factor. For the four-gluon amplitude,~${P_4 = F^4}$ where~$F$ is the linearized field strength. In tree-level open superstring theory, the color-stripped amplitude which describes the same process is,
\begin{align}
\label{eq:open}
    \cA_{\text{open}}
    &=
    P_4 \, \cA_{\text{Ven}}
\end{align}
where,
\begin{align}
\label{eq:VenIP}
    \cA_{\text{Ven}}(s,t)
    &=
    \frac{ \Gamma(-s) \Gamma(-t) }{ \Gamma(1-s-t) }
    =
    \frac{1}{st \mathstrut}
    \prod_{n \geq 1}
    \frac{ \big( 1 - \frac{ s+t \mathstrut}{ n \mathstrut } \big) }
         { \big( 1 - \frac{ s \mathstrut }{ n \mathstrut } \big)
           \big( 1 - \frac{ t \mathstrut }{ n \mathstrut } \big) }
\end{align}
Like the field theory factor~$\frac{1}{st}$, the Veneziano amplitude is symmetric in~$(s,t)$ and is a meromorphic function with simple poles only.

\subsection{Virasoro}

The Virasoro amplitude~$\cA_{\text{Vir}}$ describes the scattering of four closed strings and is a UV-completion of maximal supergravity. The tree-level field theory amplitude which describes the scattering of any four massless particles in the supergravity multiplet is,
\begin{align}
\label{eq:SUGRA}
    \cA_{\text{SG}}
    &=
    P_8 \Big( {-\frac{1}{stu}} \Big)
\end{align}
where~$P_8= \cO(s,t,u)^4$ is a kinematic pre-factor. For the four-graviton amplitude,~${P_8 = R^4}$ where~$R$ is the linearized Riemann curvature. In tree-level closed superstring theory, the amplitude which describes the same process is,
\begin{align}
\label{eq:closed}
    \cA_{\text{closed}}
    &=
    P_8 \, \cA_{\text{Vir}}
\end{align}
where,
\begin{align}
\label{eq:VirIP}
    \cA_{\text{Vir}}(s,t,u)
    &=
    \frac{ \Gamma(-s) \Gamma(-t) \Gamma(-u) }
         { \Gamma(1+s) \Gamma(1+t) \Gamma(1+u) }
    =
    - \frac{1}{stu \mathstrut}
    \prod_{n \geq 1}
    \frac{ \big( 1 + \frac{ st+tu+us \mathstrut }{ n^2 \mathstrut }
                   + \frac{ stu \mathstrut }{ n^3 \mathstrut } \big) }
         { \big( 1 - \frac{ s \mathstrut }{ n \mathstrut } \big)
           \big( 1 - \frac{ t \mathstrut }{ n \mathstrut } \big)
           \big( 1 - \frac{ u \mathstrut }{ n \mathstrut } \big) }
\end{align}
Like the field theory factor~$-\frac{1}{stu}$, the Virasoro amplitude is symmetric in~$(s,t,u)$ and is a meromorphic function with simple poles only.

\subsection{Coon}
\label{sec:Coon}

The Coon amplitude~$\cA_{q}$ is a generalization of the Veneziano amplitude with a real-valued deformation parameter~${q \geq 0}$. This deformation moves the poles of the Veneziano amplitude from the integers to the $q$-integers,
\begin{align}
    [n]_q
    =
    \frac{1-q^n}{1-q^{\phantom{n}}}
    \quad
    \xrightarrow[q \to 1]{}
    \quad
    n
\end{align}
The Coon amplitude may be written as an infinite product with a $q$-dependent pre-factor,\footnote{A more natural expression for the Coon amplitude may be given in terms of a special function called the $q$-deformed gamma function~\cite{Geiser:2022icl}.}
\begin{align}
\label{eq:CoonIP}
    \cA_{q}(s,t)
    &=
    \bigg\{
    q^{ \frac{ \ln ( 1+(q-1)s ) }{ \ln q }
        \frac{ \ln ( 1+(q-1)t ) }{ \ln q } }
    \, \Theta(1-q)
    + \Theta(q-1)
    \bigg\}
\no \\[1ex]
    & \quad
    \times
    \frac{1}{st \mathstrut}
    \prod_{n \geq 1}
    \frac{ \big( 1 - \frac{ s+t \mathstrut }{ [n]_q } 
                   + (1-q) \frac{ st \mathstrut }{ [n]_q } \big) }
        { \big( 1 - \frac{ s\phantom{_q} \mathstrut }{ [n]_q } \big)
          \big( 1 - \frac{ t\phantom{_q} \mathstrut }{ [n]_q } \big) }
\end{align}
where the step function is defined by $\Theta(x \geq 0)=1$ and $\Theta(x<0)=0$. For~${0 < q < 1}$, the poles tend to an accumulation point at~$\frac{1}{1-q}$. For~${q \geq 1}$, the poles tend to infinity. In the limits~${q \to 0}$ and~${q \to 1}$, the Coon amplitude reproduces the field theory factor and the Veneziano amplitude, respectively,
\begin{align}
    \cA_{q}(s,t)
    \quad
    & \xrightarrow[q \to 0]{}
    \quad
    \mask{\cA_{\text{Ven}}(s,t)}{\frac{1}{st}}
\no \\
    \cA_{q}(s,t)
    \quad
    & \xrightarrow[q \to 1]{}
    \quad
    \cA_{\text{Ven}}(s,t)
\end{align}
For all~$q \geq 0$, the Coon amplitude is symmetric in~$(s,t)$ with simple poles only, but its meromorphicity is subtle. For~${0 < q < 1}$, the pre-factor in~\eqref{eq:CoonIP} is explicitly non-meromorphic with branch cuts at~${s,t = \frac{1}{1-q}}$. This pre-factor ensures that the Coon amplitude has polynomial residues. For~${q \geq 1}$, there is no pre-factor, and the Coon amplitude is meromorphic.


\section{Infinite products and Weierstrass factorization}
\label{sec:infprod}

As we have seen in~\eqref{eq:VenIP}, \eqref{eq:VirIP}, and~\eqref{eq:CoonIP}, the Veneziano, Virasoro, and Coon amplitudes each have an infinite product form. Hence, we shall assume that more general tree-level scattering amplitudes with an infinite sequence of simple poles may be similarly written as infinite products. To motivate our ansatz for these generalized Veneziano and generalized Virasoro amplitudes, we shall first review some complex analysis.

\subsection{Some complex analysis}

Let ${f : U \to \bbC}$ be a function of one complex variable~$z$ on an open set ${U \subset \bbC}$. We first recall some standard definitions from single-variable complex analysis~\cite{Fischer2012}.

\begin{itemize}
\item
$f(z)$ is \textit{complex differentiable} at a point~${z_0 \in U}$ if $\smash{{\lim_{z\to z_0}} \frac{ f(z)-f(z_0) }{ z-z_0 }}$ exists.

\item
$f(z)$ is \textit{holomorphic} on~$U$ if it is complex differentiable on~$U$.

\item
$f(z)$ is \textit{meromorphic} on~$U$ if it is holomorphic on~$U$ except for a set of isolated points.

\item
$f(z)$ is \textit{entire} if it is holomorphic on the full complex plane.

\item
$f(z)$ is \textit{complex analytic} on~$U$ if for every~${z_0 \in U}$ it can be written as a convergent power series $f(z) = \sum_{n=0}^\infty a_n \, (z-z_0)^n$ with ${a_n \in \bbC}$.
\end{itemize}

It is a fundamental theorem of single-variable complex analysis that holomorphicity is equivalent to complex analyticity, but with tree-level scattering amplitudes in mind, we will be most interested in meromorphic functions.

\sm

A meromorphic function~$f(z)$ may always be written as the ratio of two holomorphic functions and is characterized by its (possibly infinite) sequence of zeros~$\zeta_n$ and poles~$\lam_n$ (counted with multiplicity). If these sequences are finite, then~$f(z)$ can be written as,
\begin{equation}
\label{eq:FiniteFact}
    f(z)
    =
    z^m \,
    e^{g(z)} \,
    \frac{ \prod_{n} ( 1-z/\mask{\lam_n}{\zeta_n} ) }
         { \prod_{n} ( 1-z/\lam_n ) }
\end{equation}
where $|m| \in \bbN$ is the order of the zero or pole at~${z=0}$, $g(z)$ is an entire function (so that~$e^{g(z)}$ has no zeros or poles), and the two finite products run over the non-zero zeros and poles. We note that each numerator and denominator factor is separately linear in the variable~$z$. This factorization is a consequence of the fundamental theorem of algebra.

\sm

If~$f(z)$ is meromorphic but with an infinite number of zeros and poles, it will admit a similar factorization. In this case, we may formally combine the two finite products in~\eqref{eq:FiniteFact} into one infinite product,
\begin{align}
\label{eq:FormalIP}
    f(z)
    =
    z^m \,
    e^{g(z)} \,
    \prod_{n} 
    \frac{ ( 1-z/\mask{\lam_n}{\zeta_n} ) }
         { (1-z/\lam_n) }
\end{align}
which converges if the zeros~$\zeta_n$ and poles~$\lam_n$ obey,
\begin{equation}\label{eq:1Converge}
    \sum_n
    \Big|
    \frac{1}{\zeta_n}
    - \frac{1}{\lam_n}
    \Big|
    <
    \infty
\end{equation}
However, the formal product~\eqref{eq:FormalIP} need not converge.

\sm

A convergent product representation of any function~$f(z)$ which is meromorphic on the full complex plane is given by the Weierstrass factorization theorem~\cite{Fischer2012}. To ensure that this (possibly infinite) product converges, it is written in terms of the elementary factors~$E_\ell(z)$,
\begin{equation}
    E_\ell(z)
    =
    \begin{cases}
    (1-z)
    &
    \ell = 0
    \\
    (1-z)
    \exp
    \big(
    \frac{z}{1}
    + \frac{z^2}{2}
    + \dots
    + \frac{z^{\ell}}{\ell}
    \big)
    &
    \ell \geq 1
    \end{cases}
\end{equation}
Using the elementary factors, it is always possible to find sequences $N_n, D_n \in \bbN$ and an entire function $\tilde{g}(z)$ such that,
\begin{equation}
\label{eq:WeirFact}
    f(z)
    =
    z^m \,
    e^{\tilde{g}(z)} \,
    \prod_n
    \frac{ E_{N_n}(z/\zeta_n) }{ E_{D_n}(z/\lam_n) }
\end{equation}
where again $|m| \in \bbN$ is the order of the zero or pole at~${z=0}$. Such a factorization always exists but is not unique. For any Weierstrass factorization, the pre-factor~$e^{\tilde{g}(z)}$ will have neither zeros nor poles.

\sm

If, in fact, the formal product~\eqref{eq:FormalIP} converges, then the entire functions~$g(z)$ and~$\tilde{g}(z)$ which appear in~\eqref{eq:FormalIP} and the general Weierstrass factorization~\eqref{eq:WeirFact} are related by,
\begin{equation}
    \exp\big( g(z) - \tilde{g}(z) \big)
    =
    \exp
    \bigg\{
    \sum_{n}
    \sum_{\ell=0}^{N_n}
    \frac{ (z/\zeta_n)^\ell }{\ell}
    -
    \sum_{n}
    \sum_{\ell=0}^{D_n}
    \frac{ (z/\lam_n)^\ell }{\ell} 
    \bigg\}
\end{equation}
where the sums $\sum_n$ are over the non-zero zeros and poles.

\sm

For example, the gamma function~$\Gamma(z)$ has the following infinite product representation in Weierstrass form,
\begin{align}
    \Gamma(z)
    =
    e^{-\gamma_E z} \,
    \frac{1}{z \mathstrut} \,
    \prod_{n\geq 1}
    \frac{1}{ \big( 1 + z/n \big ) e^{-z/n} }
\end{align}
where $\gamma_E$ is the Euler-Mascheroni constant and the denominators are just the elementary factors~$E_1(-z/n)$. This infinite product clearly demonstrates the simple poles of the gamma function at the non-positive integers, but without the factors~$e^{-z/n}$ (from the elementary factors) the product would diverge.

\sm

The Weierstrass factorization theorem is no longer applicable if the zeros and poles are bounded and tend to a common limit point~${\zeta_\infty = \lam_\infty < \infty}$ because then~$f(z)$ is no longer meromorphic at ${z = \lambda_\infty}$. In this case, however, we may still factorize the function~$f(z)$ in a form analogous to~\eqref{eq:FormalIP}. Such a factorization will not need elementary factors because the elementary factors do not improve the convergence of the infinite product in the case that~${\zeta_\infty = \lam_\infty < \infty}$. Moreover, the pre-factor~$e^{g(z)}$ in this factorization will have no zeros or poles of finite order but may have essential singularities, branch points, etc.\ since in this case the function~$f(z)$ is not meromorphic at ${z = \lambda_\infty}$.

\subsection{Ansatz for infinite product amplitudes}

We shall now use the infinite product factorization described above to motivate the ansatz for our generalized Veneziano and generalized Virasoro amplitudes. While a Weierstrass factorization~\eqref{eq:WeirFact} necessarily exists for all meromorphic functions of a single complex variable, there is no analogous theorem for functions of several complex variables. Moreover, while tree-level amplitudes are expected to be meromorphic functions of the Mandelstam variables, this need not be true. The Coon amplitude with~${q<1}$ is non-meromorphic. Thus, we shall proceed without assuming meromorphicity. Instead, we shall simply write down and analyze an infinite product ansatz analogous to~\eqref{eq:FormalIP}.

\sm

We begin with the generalized Veneziano case. Our crossing symmetric tree-level generalized Veneziano amplitude~$\cA(s,t)$ should have an infinite sequence of simple poles~$\lam_n$ in both the~$s$-channel and the~$t$-channel. We shall assume that the leading poles are at~$s=0$ and~$t=0$ and that the amplitude reduces to field theory at low-energy,
\begin{align}
    \cA(s,t) = \frac{1}{st} \big( 1 + \cO(s,t) \big)
\end{align}
Without loss of generality, we assume the poles are ordered ${\lam_n > \lam_{n-1}}$ and choose ${\lam_1 = 1}$. These assumptions can always be made true by a relabeling of the poles and a choice of units. Beyond these assumptions, the poles are wholly unspecified. In addition to its poles, $\cA(s,t)$ will have an infinite sequence of $t$\nobreakdash-dependent $s$\nobreakdash-zeros~$\zeta_n(t)$ and an identical sequence of $s$\nobreakdash-dependent $t$\nobreakdash-zeros~$\zeta_n(s)$.

\sm

Ignoring issues of convergence momentarily, we shall consider the following ansatz which satisfies these constraints and resembles the infinite product representation of the Veneziano amplitude~\eqref{eq:VenIP},
\begin{align}
\label{eq:gVen}
    \cA(s,t)
    &=
    \cW(s,t) \,
    \frac{1}{st \mathstrut}
    \prod_{n \geq 1}
    \frac{ 1-A_n(s+t)+B_n st }{ (1-s/\lam_n)(1-t/\lam_n) }
\end{align}
where $A_n$ and $B_n$ are yet undetermined coefficients and the pre-factor~$\cW(s,t) = \cW(t,s)$ has neither zeros nor poles below the largest mass pole, i.e.\ for~${|s|, |t| < \lambda_\infty}$ where~$\lambda_\infty$ may be finite or infinite. The pre-factor~$\cW(s,t)$ is analogous to the pre-factor~$e^{g(z)}$ in~\eqref{eq:FormalIP} and has the low-energy behavior~$\cW(s,t) = 1 + \cO(s,t)$. We note that the numerator and denominator of the infinite product in~\eqref{eq:gVen} are both separately linear in~$s$ and~$t$ so that the zeros and poles in either channel can be written as,
\begin{align}
    \frac{1}{s \mathstrut}
    \prod_{n \geq 1}
    \frac{ \big( 1-s/\mask{\lam_n}{\zeta_n}(t) \big) }
         { (1-s/\lam_n) \phantom{(t)} }
    \qquad
    \text{or}
    \qquad
    \frac{1}{t \mathstrut}
    \prod_{n \geq 1}
    \frac{ \big( 1-t/\mask{\lam_n}{\zeta_n}(s) \big) }
         { (1-t/\lam_n) \phantom{(t)} }
\end{align}
with the zeros given by,
\begin{align}
    \zeta_n(x)
    =
    \frac{ \mask{A_n}{1} - A_n x }{ A_n - B_n x }
\end{align}
In this form, the amplitude resembles the Weierstrass factorization~\eqref{eq:WeirFact} but without the elementary factors. The formal product in~\eqref{eq:gVen} converges if the coefficients~$A_n$ and~$B_n$ and the poles~$\lam_n$ obey,
\begin{align}
\label{eq:gVenConv}
    \sum_{n \geq 1}
    \Big| A_n - \frac{1}{\lam_n} \Big|
    &<
    \infty
    &
    \sum_{n \geq 1}
    \Big| B_n - \frac{1}{\lam_n^2} \Big|
    &<
    \infty
\end{align}
We shall return to this ansatz in~\autoref{sec:Ven}.

\sm

We now consider the generalized Virasoro case. Our crossing symmetric tree-level generalized Virasoro amplitude~$\cA(s,t,u)$ should have an infinite sequence of simple poles~$\lam_n$ in the~$s$-channel, $t$-channel, and~$u$-channel. We shall again assume that the leading poles are at $s=0$, $t=0$, and $u=0$ and that the amplitude reduces to field theory at low-energy,
\begin{align}
    \cA(s,t,u) = -\frac{1}{stu} \big( 1 + \cO(s,t,u) \big)
\end{align}
Without loss of generality, we assume the poles are ordered ${\lam_n > \lam_{n-1}}$ and choose ${\lam_1 = 1}$. Beyond these assumptions, the poles are again wholly unspecified.

\sm

Again momentarily ignoring issues of convergence, we shall consider the following ansatz which satisfies these constraints and resembles the infinite product representation of the Virasoro amplitude~\eqref{eq:VirIP},
\begin{align}
\label{eq:gVir}
    \cA(s,t,u)
    &=
    \cW(s,t,u) \,
    \Big( {-\frac{1}{stu \mathstrut}} \Big)
    \prod_{n \geq 1}
    \frac{ 1+A_n(st+tu+us)-B_n stu }
         { (1-s/\lam_n)(1-t/\lam_n)(1-u/\lam_n) }
\end{align}
where $A_n$ and $B_n$ are yet undetermined coefficients and the $(s,t,u)$-symmetric pre-factor ${\cW(s,t,u)}$ has neither zeros nor poles below the largest mass pole, i.e.\ for~${|s|, |t|, |u| < \lambda_\infty}$ where~$\lambda_\infty$ may again be finite or infinite. As before, the pre-factor~$\cW(s,t,u)$ is analogous to the pre-factor~$e^{g(z)}$ in~\eqref{eq:FormalIP} and has the low-energy behavior~$\cW(s,t,u) = 1 + \cO(s,t,u)$. We note that the numerator and denominator of the infinite product in~\eqref{eq:gVir} are both separately linear in~$s$, $t$, and~$u$. Moreover, there is no term proportional to~${s+t+u}$ in the numerator because this combination vanishes on-shell for massless external states. The formal product in~\eqref{eq:gVir} converges if the coefficients $A_n$ and $B_n$ and the poles $\lam_n$ obey,
\begin{align}
\label{eq:gVirConv}
    \sum_{n \geq 1}
    \Big| A_n - \frac{1}{\lam_n^2} \Big|
    &<
    \infty
    &
    \sum_{n \geq 1}
    \Big| B_n - \frac{1}{\lam_n^3} \Big|
    &<
    \infty
\end{align}
We shall return to this ansatz in~\autoref{sec:Vir}.

\sm

In both the generalized Veneziano and generalized Virasoro case, demanding that the $t$-channel poles cancel on each $s$-channel pole will enforce strong constraints on the undetermined coefficients~$A_n$ and~$B_n$ as well as the poles~$\lam_n$. In the following two sections, we shall analyze these constraints in detail.


\section{Generalized Veneziano amplitudes}
\label{sec:Ven}

In this section, we shall systematically analyze our infinite product ansatz~\eqref{eq:gVen} for the generalized Veneziano amplitude.

\subsection{Veneziano truncation}

We first recall the infinite product form~\eqref{eq:VenIP} of the Veneziano amplitude, which has simple poles at each non-negative integer. The residue of the massless $s$\nobreakdash-channel pole is~${1/t}$, and the residue of each massive pole at~${s=N}$ is a polynomial of degree\nobreakdash-${(N-1)}$ in~$t$. The Veneziano amplitude achieves these residues because on each $s$-pole, its zeros cancel the $t$\nobreakdash-poles, leaving a finite polynomial in~$t$. This cancellation can be described in terms of the numerator factors,
\begin{align}
    \cN_n(s,t)
    =
    1-(s+t)/n
\end{align}
When~$s=N$, each numerator factor ${\cN_{N+n}(N,t)=\frac{n}{N+n}(1-t/n)}$ cancels the $t$-channel pole from the factor~${(1-t/n)^{-1}}$, and the infinite product truncates. In short, the condition,
\begin{align}
\label{eq:VenTrunc}
    \cN_{N+n}(N,n) = 0
\end{align}
ensures that the Veneziano amplitude has polynomial residues.

\subsection{Generalized Veneziano truncation}

We now return to our generalized Veneziano ansatz~\eqref{eq:gVen}. We shall demand that the zeros and poles of this amplitude cancel in a similar fashion as those of the Veneziano amplitude. We first demand that the residue at~${s=0}$ is~${1/t}$ so that the amplitude reproduces the massless spectrum of super Yang-Mills analogously to the Venziano amplitude,
\begin{align}
    \Res_{s=0}
    \cA(s,t)
    =
    \frac{1}{t}
    \qquad
    \implies
    \qquad
    \cW(0,t) \,
    \prod_{n \geq 1}
    \frac{ 1- \mask{t/\lam_n}{A_n t} }
         { 1 - t/\lam_n }
    = 1
\end{align}
which implies that $\cW(0,t) = 1$ and~$A_n = 1/\lam_n$ since~$\cW(s,t)$ has neither zeros nor poles. In other words, the coefficients~$A_n$ are determined by the poles~$\lambda_n$.

\sm

Next, in analogy with the truncation condition for the Veneziano amplitude~\eqref{eq:VenTrunc}, we demand that the generalized numerator factor,
\begin{align}
    {\cN_n(s,t) = 1 - A_n (s+t) + B_n st}
\end{align}
obeys the generalized truncation condition,
\begin{align}
\label{eq:gVenTrunc}
    \cN_{N+n}(\lambda_N,\lambda_n) = 0
\end{align}
so that ${\cN_{N+n}(\lam_N,t) \propto (1-t/\lam_n)}$ and the infinite sequence of $t$-channel poles cancels on each $s$-channel pole.\footnote{At this point, we are no longer considering the most general possible infinite product amplitude but are instead working in close analogy with the Veneziano amplitude. A more general truncation condition,  ${\cN_{N+n+\alpha}(\lambda_N,\lambda_n) = 0}$ for some positive integer $\alpha$, is considered in~\cite{Cheung:2022mkw}.} This truncation condition determines the coefficients~$B_n$ in terms of the poles~$\lam_n$,
\begin{align}
    B_n
    &=
    \frac{\lam_{k}+\lam_{n-k}-\lam_n}{\lam_n\lam_{n-k}\lam_k}
    &
    k
    &=
    1,2,\dots,n-1
\end{align}
For fixed ${n \geq 2}$, both~$k$ and~${k'=n-k}$ yield the same equation for~$B_n$ so that there are~$\floor{\frac{n}{2}}$ independent equations for~$B_n$, where~$\floor{x}$ is the floor function. The coefficient~$B_1$ is left undetermined, the coefficients~$B_2$ and~$B_3$ are uniquely determined, and the coefficients~$B_n$ with~${n\geq 4}$ are all over-determined.

\sm

This over-determination of the~$B_n$ highly constrains the poles. Any sequence of poles~$\lam_n$ must leave the following combination independent of~$k$ for all~${n \geq 2}$,
\begin{align}
\label{eq:VenConstraints}
    \Lambda_n(k)
    &=
    \frac{\lam_{k}+\lam_{n-k}-\lam_n}{\lam_{n-k}\lam_k}
\end{align}
We shall refer to these equations as the generalized Veneziano amplitude constraints. The Veneziano solution~${\lam_n=n}$ (i.e.\ the string theory spectrum) solves these constraints with~${\Lambda_n(k)=0}$ for all~$n$ and~$k$. We shall search for other, more general sequences of poles~$\lam_n$ which solve the generalized Veneziano amplitude constraints.

\subsection{Generalized Veneziano amplitude constraints}

Since~$\Lambda_n(k)$ must be independent of~$k$, we may fix~$n \geq 2$ and choose two distinct values of~$(k,k')$ in the appropriate range to find,
\begin{align}
\label{eq:VenReckk'}
    \Lambda_n(k)
    =
    \Lambda_n(k')
    \qquad
    \implies
    \qquad
    \frac{\lam_{k}+\lam_{n-k}-\lam_n}{\lam_{n-k}\lam_{k}}
    =
    \frac{\lam_{k'}+\lam_{n-k'}-\lam_n}{\lam_{n-k'}\lam_{k'}}
\end{align}
This equation is a non-linear recursion relation for the poles~$\lam_n$ of order~$\text{max}(k,k')$ which determines all the~$\lam_n$ with~${n > \text{max}(k,k')}$ in terms of the lower~$\lam_n$ (except for ${n=k+k'}$, in which case the equation is vacuous). Because we are free to choose~$(k,k')$ within the appropriate range, the poles~$\lam_n$ must solve an infinite set of these non-linear recursion relations. This system is highly constrained, and there is no guarantee that a general solution (other than the Veneziano solution) exists!

\sm

It turns out, however, that from~\eqref{eq:VenReckk'} we can derive a simple first-order recursion relation which determines all the poles~${\lam_{n}}$ with~${n \geq 4}$ in terms of~$\lambda_1$, $\lambda_2$, and~$\lambda_3$. We consider the following three equations for fixed~${n \geq 4}$,
\begin{align}
    \Lambda_n(1)
    &=
    \Lambda_n(2)
    &
    \Lambda_n(1)
    &=
    \Lambda_n(3)
    &
    \Lambda_{n-1}(1)
    &=
    \Lambda_{n-1}(2)
\end{align}
These three equations include the poles~$\lam_1$, $\lam_2$, $\lam_3$, $\lam_{n-3}$, $\lam_{n-2}$, $\lam_{n-1}$, and~$\lam_n$, but we may eliminate~$\lam_{n-3}$ and~$\lam_{n-2}$ to find the following first order recursion relation for~$\lam_n$ in terms of only~$\lam_1$, $\lam_2$, $\lam_3$, and~$\lam_{n-1}$,
\begin{align}
\label{eq:Riccati}
    \lam_n
    &=
    \frac{a\lam_{n-1}+b}{c\lam_{n-1}+d}
\end{align}
where the coefficients $a$, $b$, $c$, and $d$ are given by,
\begin{align}
    a
    &=
    \lam_2(1-2\lam_3+\lam_2\lam_3)
    =
    (1+x)(x^2+xy-y)
    \no \\
    b
    &=
    \mask{\lam_2(1-2\lam_3+\lam_2\lam_3)}{\lam_2(\lam_3-\lam_2)}
    =
    (1+x)y
    >
    0
    \no \\
    c
    &=
    \mask{\lam_2(1-2\lam_3+\lam_2\lam_3)}{1+\lam_2^2-\lam_2-\lam_3}
    =
    x^2-y
    \no \\
    d
    &=
    \mask{\lam_2(1-2\lam_3+\lam_2\lam_3)}{\lam_2(\lam_3-\lam_2)}
    =
    (1+x)y
    >
    0
\end{align}
Here we have defined the positive numbers~${x=\lam_2-\lam_1=\lam_2-1>0}$ and~${y=\lam_3-\lam_2>0}$, using the fact that the poles~${\lam_n > \lam_{n-1}}$ are ordered. The recursion relation~\eqref{eq:Riccati} was derived for~${n \geq 4}$ but is in fact true for all~${n \geq 1}$ if we define ${\lam_0 = 0}$. For $n=1,2,3$,~\eqref{eq:Riccati} is only vacuously true and does not determine~$\lam_1$, $\lam_2$, or~$\lam_3$. The choice~${\lam_1 = 1}$ simply sets our units, and the free parameters~$\lam_2$ and~$\lam_3$ (or equivalently~$x$ and~$y$), define a two-parameter space of possible solutions in the region~${x, y > 0}$. The string spectrum~${\lam_n = n}$ is at the point ${x=y=1}$ of this two-parameter space.

\subsection{Solving the Riccati relation}
\label{subsec:Ricc}

The recursion relation~\eqref{eq:Riccati} is known as the Riccati recursion relation with constant coefficients, and its solutions are well known. An exhaustive study of non-linear recursion relations of this kind may be found in~\cite{kocic1993}.

\sm

Although the Riccati recursion relation~\eqref{eq:Riccati} is generally non-linear, there is a curve ${c=x^2-y=0}$ in parameter space where it becomes linear,
\begin{equation}
    \lam_n
    =
    x \lam_{n-1}+1
\end{equation}
and yields the Coon spectrum (relabeling~${x\to q}$),
\begin{equation}
\label{eq:qSoln}
    \lam_n
    =
    \frac{1-q^n}{1-q^{\phantom{n}}}
\end{equation}
The Coon spectrum reproduces the string spectrum at~${q = 1}$ and accounts for all the spectra reviewed in~\autoref{sec:review}. For~${q>1}$, the poles grow exponentially, and for~${0<q<1}$, they monotonically accumulate to the limit point~${\lam_{\infty}=\frac{1}{1-q}}$.

\sm

Beyond these well-studied solutions, there is, however, a much larger space of solutions to~\eqref{eq:Riccati} with~${c=x^2-y\neq 0}$. In this case, the non-linear first-order recursion relation~\eqref{eq:Riccati} can be reduced to a linear second-order recursion relation using the following change of variables,
\begin{equation}
    c\lam_n+d
    =
    (a+d)\frac{z_{n+1}}{z_n}
\end{equation}
with the boundary condition~${z_0 = 1}$. Substituting this expression into~\eqref{eq:Riccati}, we find,
\begin{align}
\label{eq:VenlinearRec}
    z_{n+2}
    -
    z_{n+1}
    + R \, z_n
    =
    0
\end{align}
with the positive coefficient~$R$ given by,
\begin{align}
    R
    =
    \frac{ad-bc}{(a+d)^2}
    =
    \frac{y}{(1+x)(x+y)}
    >
    0
\end{align}
The solutions of this linear recursion relation are governed by the quadratic equation,
\begin{equation}
\label{eq:polynomial}
    r^2-r+R=0
\end{equation}
whose roots are,
\begin{align}
\label{eq:VenRoots}
    r_{\pm}
    =
    \frac{1\pm\sqrt{1-4R}}{2}
\end{align}
We shall separately analyze the cases~${R \neq \frac{1}{4}}$ and~${R = \frac{1}{4}}$.

\subsubsection{The case \texorpdfstring{${R \neq \frac{1}{4}}$}{}}

If~${R \neq \frac{1}{4}}$, then the roots~${r_+ \neq r_-}$ are distinct and~$z_n$ is given by,
\begin{align}
    z_n
    =
    \frac{ \mask{r_+}{z_1}-r_- }{ r_+-r_- }
    \, r_+^n
    +
    \frac{ r_+- \mask{r_-}{z_1} }{ r_+-r_- }
    \, r_-^n
\end{align}
Subsequently,~$\lam_n$ is given by,
\begin{align}
\label{eq:pSoln}
    \lam_n
    =
    \frac{(1+x)(1-p^n)}
         {(1-xp)-(1-x/p)p^n}
\end{align}
with~${p = r_- / r_+}$ so that~${|p| \leq 1}$. We shall refer to these solutions as~$p$-type spectra. These spectra were first identified in~\cite{Coon:1969yw} and were later called M\"obius trajectories in~\cite{Fairlie:1994ad}. Our parametrization in terms of~$p$ is novel and can be clearly related to the first three mass levels through the parameters~$x$ and~$y$ since,
\begin{align}
    p
    =
    p(x,y)
    =
    \frac{1-\sqrt{1-{4 y/}{(1+x)(x+y)}}}
         {1+\sqrt{1-{4 y}/{(1+x)(x+y)}}}
\end{align}
When~${p = x}$ or ${p = x^{-1}}$, these spectra reduce to the Coon solution~\eqref{eq:qSoln} with~${q = p < 1}$ or~${q = p^{-1} > 1}$, respectively.

\sm

The expression~\eqref{eq:pSoln} solves the generalized Veneziano amplitude constraints~\eqref{eq:VenConstraints} for all~${x,y>0}$, but the resultant~$\lam_n$ will not necessarily be monotonically ordered and positive. We shall now determine the values of~$x$ and~$y$ which yield a monotonically increasing sequence of poles~$\lam_n$.

\sm

We first note that ${R > \frac{1}{4}}$ implies that the parameter~${p = r_-/r_+ = e^{i\phi}}$ is a phase so that the~$\lam_n$ are periodic as a function of~$n$. These periodic solutions always produce negative (and thus unphysical) $\lam_n$. The condition~${R > \frac{1}{4}}$ is equivalent to $y > \frac{x(1+x)}{3-x}$, so this region of parameter space is ruled out.

\sm

We now consider ${0 < R < \frac{1}{4}}$ which corresponds to ${0 < y < \frac{x(1+x)}{3-x}}$. In this case, the roots~$r_{\pm}$ are real and positive, so the parameter~$p$ is in the range~${0 < p < 1}$. To determine whether the~$\lam_n$ increase monotonically in this region, we shall momentarily treat~$n$ as a continuous variable so that~${\lam_n \to \lam(n)}$ is an analytic function of~$n$ with a discrete set of singularities at the points~${n=n_*}$ on the complex $n$-plane,
\begin{align}
\label{eq:Singularity}
    n_*
    =
    \frac{ \ln \big( \frac{1-xp \phantom{/} }{1-x/p} \big) }
         { \ln p }
    - 
    \frac{2 \pi i k}{ \ln p }
\end{align}
with~${k \in \bbZ}$. Since $\frac{d}{dn} \lam(n) > 0$ for all real~${n \geq 0}$, the function $\lam(n)$ can only fail to be monotonic if there is a singularity~${n_* > 0}$ on the real $n$-axis such that ${\lim_{n \to n_*^{\mp}} \lam(n) = \pm\infty}$. From~\eqref{eq:Singularity}, we see that there is at most one such singularity with ${k=0}$, which occurs if and only if,
\begin{equation}
\label{eq:condition}
    0 < \frac{ 1-xp \phantom{/} }{ 1-x/p } < 1
\end{equation}
We first suppose that~\eqref{eq:condition} is satisfied with~${1-xp > 0}$ and~${1-x/p > 0}$, which then implies ${1-xp < 1-x/p}$ and thus~${p>1}$. Since ${0<p<1}$, we must instead have~${1-xp<0}$ and~${1-x/p<0}$. To proceed, we shall separately consider the cases~${0 < x < 1}$ and~${x \geq 1}$.
\begin{itemize}
\item
For~${0<x<1}$, it is not possible to fulfill the condition~${1-xp<0}$, so the whole region corresponding to~${0 < R < \frac{1}{4}}$ and~${0<x<1}$ yields monotonically increasing and positive~$\lam_n$. 
\item
For~${x\geq 1}$, the condition ${1-x/p<0}$ is always satisfied, but~${1-xp<0}$ implies,
\begin{align}
    x
    >
    \frac{1}{p(x,y)}
    \quad
    \implies
    \quad
    \bigg( \frac{x-1}{x+1} \bigg)^2
    >
    \frac{x(1+x)-(3-x)y}{(1+x)(x+y)}
     \quad
    \implies
    \quad
    y
    >
    x^2
\end{align}
Therefore, when either~${1 \leq x < 3}$ and ${x^2 < y < \frac{x(1+x)}{3-x}}$ or when ${x \geq 3}$ and ${y > x^2}$, the function~$\lambda(n)$ has a singularity at finite~${n = n_* > 0}$ and is not monotonic. Moreover, in this region, the limit point~${\lambda_\infty = \frac{1+x \phantom{p}}{1-xp} < 0}$ is negative and thus non-physical.
\end{itemize}

\subsubsection{The case \texorpdfstring{${R = \frac{1}{4}}$}{}}

If~${R = \frac{1}{4}}$, then the roots~${r_+ = r_-}$ are equal and~$z_n$ is given by,
\begin{equation}
    z_n
    =
    \frac{1}{2^n}
    \Big(
    1+\frac{1-x}{2x}n
    \Big)
\end{equation}
Subsequently,~$\lam_n$ is given by,
\begin{equation}
\label{eq:RSoln}
    \lam_n
    =
    \frac{(1+x)n}{2x+(1-x)n}
\end{equation}
We shall refer to these solutions as~$r$-type spectra (where the $r$ is for rational). The expression~\eqref{eq:RSoln} solves the generalized Veneziano amplitude constraints~\eqref{eq:VenConstraints} for all~${x>0}$, but the resultant~$\lam_n$ will only be monotonically ordered and positive for~${0 < x \leq 1}$. When~${x>1}$, the limit point~${\lambda_\infty = \frac{1+x}{1-x} < 0}$ is negative and thus non-physical. When ${x = 1}$, this solution reduces to the string spectrum~${\lam_n = n}$. Notably, these $r$-type solutions were not identified in the previous literature~\cite{Coon:1969yw, Fairlie:1994ad}.

\subsubsection{Summary}

\begin{figure}

\centering

\begin{tikzpicture}

\begin{axis}
[
    width = 1\textwidth,
    height = 0.7\textwidth,
    xmin = 0, xmax = 1.5,
    ymin = 0, ymax = 1.5,
    xtick = {0,1},
    ytick = {1},
    extra y ticks = {0},
    extra y tick labels = \empty,
    xlabel = {$x$},
    ylabel = {$y$},
    x label style = {at={(current axis.right of origin)}, anchor=west},
    y label style = {at={(current axis.north west)}, rotate=-90, anchor=south},
    axis lines = left,
    legend cell align = {left},
    legend pos = north west,
    legend entries =
    {
        \, string spectrum $x=y=1$,
        \, \makebox[\widthof{string spectrum}][l]{Coon spectra} \smash{$y=x^2$},
        \, \makebox[\widthof{string spectrum}][l]{$r$-type spectra} \smash{$y=x(1+x)/(3-x)$},
        \, $p$-type spectra,
        \, unphysical spectra
    }
]

\addlegendimage{only marks, mark size = 2pt, fill = green}
\addlegendimage{ultra thick, color = black}    
\addlegendimage{ultra thick, color = black, densely dashed}
\addlegendimage{only marks, mark = square*, mark size = 4pt, color = yellow!40}
\addlegendimage{only marks, mark = square*, mark size = 4pt, color = red!30}

\path[name path = xtop] (0,1.5) -- (1.5,1.5);
\path[name path = xbottom] (0,0) -- (1.5,0);

\addplot
[
    name path = c,
    domain = 0:1.5, 
    ultra thick,
	black,
    smooth
]
{x*x};

\addplot
[
    name path = r,
    domain = 0:1, 
    ultra thick,
	black,
    densely dashed,
    smooth
]
{x*(1+x)/(3-x)};
  
\addplot
[
    only marks,
    mark size = 4pt,
    fill = green
]
coordinates {(1,1)};

\addplot[yellow!40]
fill between
[
    of = r and c,
    soft clip = {domain = 0:1},
    on layer = axis background
];

\addplot[yellow!40]
fill between
[
    of = c and xbottom,
    soft clip = {domain = 0:1.2247},
    on layer = axis background
];

\addplot[yellow!40]
fill between
[
    of = xtop and xbottom,
    soft clip = {domain = 1.224:1.5},
    on layer = axis background
];

\addplot[red!30]
fill between
[
    of = r and xtop,
    soft clip = {domain = 0:1},
    on layer = axis background
];

\addplot[red!30]
fill between
[
    of = c and xtop,
    soft clip = {domain = 0.99:1.2247},
    on layer = axis background
];

\end{axis}

\end{tikzpicture}

\caption{The two-parameter space of solutions to the generalized Veneziano amplitude constraints. The point ${x=y=1}$ corresponds to the string spectrum. The solid black line corresponds to the one-parameter subspace of Coon spectra. The dashed black line corresponds to the one-parameter subspace of $r$\nobreakdash-type spectra. The yellow region corresponds to the two-parameter subspace of $p$\nobreakdash-type spectra. The red region corresponds to unphysical spectra with negative mass squared.}

\label{fig:Ven1}

\end{figure}
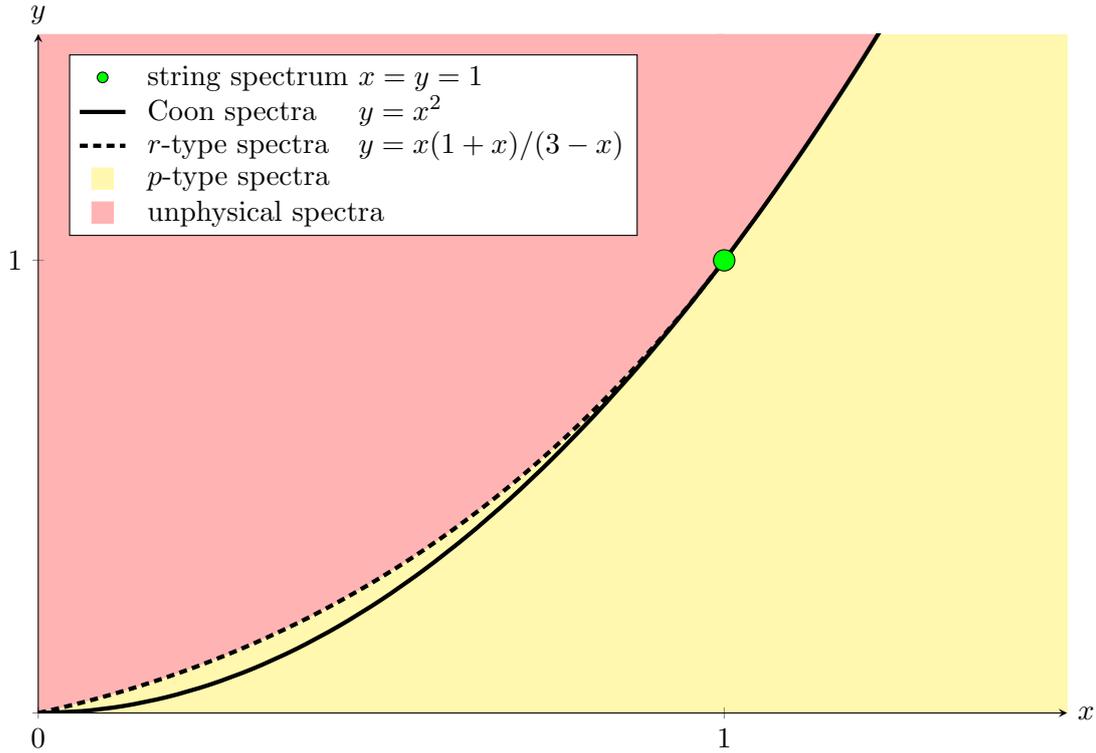

We have now fully classified all the monotonically ordered and positive solutions of the generalized Veneziano amplitude constraints~\eqref{eq:VenConstraints}. These solutions exist in the region of the $xy$\nobreakdash-plane defined by,
\begin{align}
    \Big\{
    0 < x < 1
    \,\, , \,\,
    0 < y \leq \tfrac{x(1+x)}{3-x}
    \Big\}
    \cup
    \Big\{
    1 \leq x
    \,\, , \,\,
    0 < y \leq x^2
    \Big\}
\end{align}
where again ${x = \lambda_2-\lambda_1 > 0}$ and ${y = \lambda_3 - \lambda_2 > 0}$ are positive parameters which determine the second and third masses. This region and the various solutions are shown in~\autoref{fig:Ven1}. Notably, all the non-monotonically-ordered solutions to the Riccati equation~\eqref{eq:Riccati}, i.e.\ the points within the excluded regions of parameter space, yield negative~$\lam_n$ and are unphysical.

\sm

For completeness, we shall rewrite all the solutions and the ranges of their parameters. The Coon spectra have one free parameter and are given by,
\begin{align}
\label{eq:VenqSolns}
    \lam_n
    &=
    \frac{1-q^n}{1-q^{\phantom{n}}}
    &
    0 &< q < \infty
\intertext{where $q$ is related to the parameters~$x$ and~$y$ by~${x=q}$ and~${y=q^2}$. The $p$\nobreakdash-type spectra have two free parameters and are given by,}
\label{eq:VenpSolns}
    \lam_n
    &=
    \frac{(1+x)(1-p^n)}{(1-xp)-(1-x/p)p^n}
    &
    0 &< x < \infty
    &
    0 &< p < \min(1, x^{-1})
\no \\
    &&&&&
    \phantom{{}<{}}
    p \neq x, x^{-1}
\intertext{where we have excluded~${p = x, x^{-1}}$ to avoid double-counting the Coon spectra. Finally, the $r$\nobreakdash-type spectra have one free parameter and are given by,}
\label{eq:VenRSolns}
    \lam_n
    &=
    \frac{(1+x)n}{2x+(1-x)n}
    &
    0 &< x < 1
\end{align}
The string spectrum~${\lam_n=n}$ is located at ${x = y = 1}$ in parameter space and can be obtained by taking various limits of each of these solutions. All of these spectra have a finite accumulation point $\lambda_\infty$, except for the Coon spectra~\eqref{eq:VenqSolns} with~${q \geq 1}$ (which includes the string spectrum at~${q=1}$). 

\sm

Although we derived these solutions from the Riccati relation~\eqref{eq:Riccati}, they do in fact satisfy the full generalized Veneziano amplitude constraints~\eqref{eq:VenConstraints}. For each case, we may compute~$\Lambda_n(k)$ and verify that it is independent of $k$. Since any solution of the generalized Veneziano amplitude constraints~\eqref{eq:VenConstraints} necessarily satisfies the Riccati relation~\eqref{eq:Riccati}, we have thus fully solved~\eqref{eq:VenConstraints}. Explicitly, the~$\Lambda_n(k)$ are given by,
\begin{align}
    \text{Coon} :
    \qquad
    \Lambda_n(k)
    &=
    1-q
    \vphantom{\frac{1-x}{1+x}}
\no \\[1ex]
    p\text{-type} :
    \qquad
    \Lambda_n(k)
    &=
    \frac{1}{(1+x)}
    \cdot
    \frac{(1-xp)^2-(1-x/p)^2p^n}
         {(1-xp)^{\phantom{2}}-(1-x/p)^{\phantom{2}}p^n}
\no \\[1ex]
    r\text{-type} :
    \qquad
    \Lambda_n(k)
    &=
    \frac{1-x}{1+x}
    \cdot
    \frac{4x+(1-x)n}{2x+(1-x)n}
\intertext{These three expression can be written in a universal form,}
\label{eq:Lamnk}
    \Lambda_n(k)
    &=
    \frac{1}{\lam_\infty}
    + \frac{1}{ \lam_{-\infty} }
    - \frac{ \lam_n }{ \lam_{\infty} \lam_{-\infty} }
\end{align}
where we have defined the (possibly infinite) quantities ${\lam_{\pm\infty} = \lim_{n \to \pm\infty} \lam_n}$. We have,
\begin{align}
    \text{Coon} \, (q < 1) :
    &&
    \frac{1}{\lam_\infty}
    &=
    1-q
    &
    \frac{1}{\lam_{-\infty}}
    &=
    0
\no \\[1ex]
    \text{Coon}  \, (q \geq 1) :
    &&
    \frac{1}{\lam_\infty}
    &=
    0
    &
    \frac{1}{\lam_{-\infty}}
    &=
    1-q
\no \\[1ex]
    p\text{-type} :
    &&
    \frac{1}{\lam_\infty}
    &=
    \frac{ 1-xp }{ 1+x \phantom{p} }
    &
    \frac{1}{\lam_{-\infty}}
    &=
    \frac{ 1-x/p }{ 1+x \phantom{/p} }
\no \\[1ex]
    r\text{-type} :
    &&
    \frac{1}{\lam_\infty}
    &=
    \frac{ 1-x }{ 1+x }
    &
    \frac{1}{\lam_{-\infty}}
    &=
    \frac{ 1-x }{ 1+x }
\end{align}
When $\lam_\infty$ is finite, it is of course the limit point of poles. The quantity $\lam_{-\infty}$ does not have a clear physical interpretation but is still useful to define.

\sm

Finally, we note that with~${A_n=1/\lam_n}$ and~${B_n=\Lambda_n/\lam_n}$, all of the solutions we find give convergent infinite product amplitudes~\eqref{eq:gVen} satisfying the convergence condition~\eqref{eq:gVenConv}. As we noted above, the coefficient~$B_1$ is undetermined by our constraints, but we shall choose ${B_1=\Lambda_1/\lam_1}$ to fit the pattern. This choice will not affect our subsequent analysis.

\subsection{Polynomial residues?}

We derived the spectra above from the generalized Veneziano amplitude constraints~\eqref{eq:VenConstraints}, which we in turn derived by demanding that the infinite sequence of $t$-channel poles cancels on each $s$-channel pole within our infinite product ansatz~\eqref{eq:gVen}. However, this truncation condition will not necessarily imply that our generalized Veneziano amplitudes have polynomial residues. With our explicit expressions for the poles~$\lam_n$ and the coefficients~$A_n$ and~$B_n$, we can explicitly compute the residues of~\eqref{eq:gVen}. 

\sm

We shall denote the Coon amplitudes by~$\cA_q(s,t)$, the $p$-type amplitudes by~$\cA_p(s,t)$, and the $r$-type amplitudes by~$\cA_r(s,t)$. We may then manipulate~\eqref{eq:gVen} and write each of these amplitudes in a form such that each factor in its infinite product is manifestly convergent. For the Coon amplitudes, we have,
\begin{align}
\label{eq:Aq}
    \cA_q(s,t)
    &=
    \cW_q(s,t) \,
    \frac{1}{st}
    \prod_{n\geq 1}
    \frac{ (1-\hat{q}^{n-\alpha_q(s)-\alpha_q(t)}) (1-\hat{q}^{n}) }
         { (1-\hat{q}^{n-\alpha_q(s)}) (1-\hat{q}^{n-\alpha_q(t)})}
\no \\[1ex]
    \alpha_q(s)
    &=
    \frac{ \ln \big( 1+(q-1)s \big) }{ \ln q }
\intertext{where $\hat{q} = \min(q,q^{-1})$. For the $p$-type amplitudes, we have,}
\label{eq:Ap}
    \cA_p(s,t)
    &=
    \cW_p(s,t) \,
    \frac{1}{st}
    \prod_{n\geq 1}
    \frac{ (1-p^{n-\alpha_p(s)-\alpha_p(t)}) (1-p^{n}) }
         { (1-p^{n-\alpha_p(s)}) (1-p^{n-\alpha_p(t)})}
\no \\[1ex]
    \alpha_p(s)
    &=
    \frac{ \ln \Big( \frac{ (1+x)-(1-xp)s \phantom{/} }
                          { (1+x)-(1-x/p)s } \Big) }
         { \ln p }
\intertext{Finally, for the $r$-type amplitudes, we have,}
\label{eq:Ar}
    \cA_r(s,t)
    &=
    \cW_r(s,t) \,
    \frac{1}{st}
    \prod_{n\geq 1}
    \frac{ 1 - \big( \alpha_r(s)+\alpha_r(t) \big) / n }
         { \big( 1-\alpha_r(s)/n \big)
           \big( 1-\alpha_r(t)/n \big) }
\no \\[1ex]
    \alpha_r(s)
    &=
    \frac{2xs}{1+x-(1-x)s}
\end{align}
In each case, the functions~$\alpha(s)$ are the respective amplitudes' leading Regge trajectory and obey ${\alpha(\lam_N)=N}$.

\sm

For simplicity, we have omitted the exponential factors needed to make the infinite product of each factor in~\eqref{eq:Ar} convergent. As in the infinite product representation for the Veneziano amplitude~\eqref{eq:VenIP}, these factors cancel between the numerator and denominator.

\sm

From these expressions, we may simply compute the residues at $s=\lam_N$ for $N\geq 1$. We recall that the residue of the massless pole at~${s = 0}$ is $1/t$ by construction. For the massive poles, we find the following. For the Coon amplitudes with~${q \geq 1}$, we have,
\begin{align}
    \Res_{s=\lam_N}
    \cA_q(s,t)
    &=
    \cW_q(\lam_N,t) \,
    \frac{q^N}{\lambda_N}
    \prod_{n=1}^{N-1}
    \bigg(
    \frac{q^n}{\lambda_n}
    \,
    t
    + 1
    \bigg)
\intertext{For the Coon amplitudes with~${q < 1}$, we have,}
    \Res_{s=\lam_N}
    \cA_q(s,t)
    &=
    \cW_q(\lam_N,t) \,
    \frac{q^N}{\lambda_N}
    \frac{1}{ (1-t/\lam_\infty)^N }
    \prod_{n=1}^{N-1}
    \bigg(
    \frac{q^n}{\lambda_n}
    \,
    t
    + 1
    \bigg)
\intertext{For the $p$-type amplitudes, we have,}
    \Res_{s=\lam_N}
    \cA_p(s,t)
    &=
    \cW_p(\lam_N,t) \,
    \frac{p^N}{\lambda_N}
    \frac{ x^2 (1-p^2)^2 }
         { \big[ p(1-xp)-(p-x)p^N \big]^2 }
\no \\
    & \quad
    \times
    \frac{1}{ (1-t/\lam_\infty)^N }
    \prod_{n=1}^{N-1}
    \bigg(
    \frac{ (1-xp)p^n - (1-x/p) }{(1+x)(1-p^n) }
    \,
    t
    + 1
    \bigg)
\intertext{Finally, for the $r$-type amplitudes, we have,}
    \Res_{s=\lam_N}
    \cA_r(s,t)
    &=
    \cW_r(\lam_N,t) \,
    \frac{1}{\lambda_N}
    \frac{ 4x^2 }{ \big( 2x+(1-x)N \big)^2 }
\no \\
    & \quad
    \times
    \frac{1}{ (1-t/\lam_\infty)^N }
    \prod_{n=1}^{N-1}
    \bigg(
    \frac{ 2x-(1-x)n }{ (1+x)n }
    \,
    t
    + 1
    \bigg)
\intertext{Using the quantities~$\lam_{\pm\infty}$ defined above, we may write these expressions in the following universal form,}
\label{eq:Res}
    \Res_{s=\lam_N}
    \cA(s,t)
    &=
    \cW(\lam_N,t) \,
    \frac{ 1 }{ \lam_N }
    \bigg( 1 - \frac{\lam_N}{\lam_\infty} \bigg)
    \bigg( 1 - \frac{\lam_N}{\lam_{-\infty}} \bigg)
\no \\
    & \quad
    \times
    \frac{1}{ (1-t/\lam_\infty)^N }
    \prod_{n=1}^{N-1}
    \bigg[
    \bigg(
    \frac{1}{\lambda_n}
    - \frac{1}{\lambda_{\infty}}
    - \frac{1}{\lambda_{-\infty}}
    \bigg)
    \,
    t
    + 1
    \bigg]
\end{align}
Ignoring for now the~$\cW(s,t)$ pre-factors, only the amplitudes $\cA_q(s,t)$ with ${q \geq 1}$ have polynomial residues. In other words, the amplitudes with accumulation point spectra all have non-polynomial residues! In each case, though, the non-polynomial behavior is captured by the factor~${(1-t/\lam_\infty)^{-N}}$ which multiplies a degree-$(N-1)$ polynomial in~$t$. 

\sm

This, however, is not the end of the story. It may be possible to find a pre-factor~$\cW(s,t)$ which cancels the non-polynomial factors~${(1-t/\lam_\infty)^{-N}}$ on each pole. We recall that the pre-factor obeys ${\cW(s,t) = 1 + \cO(s,t)}$. We must then require ${\cW(\lam_N,t) \propto (1-t/\lam_\infty)^{N}}$ for all~${N \geq 1}$ to cancel the non-polynomial factors in each residue. A natural guess is simply ${\cW(s,t) = (1-t/\lam_\infty)^{\alpha(s)}}$ for the appropriate Regge trajectory $\alpha(s)$.\footnote{A more general pre-factor is considered in~\cite{Cheung:2022mkw}.} In fact, any pre-factor~$\cW(s,t)$ which cancels the non-polynomial behavior on every residue must be proportional to this guess, but this guess is not generally crossing symmetric. Only for the Coon amplitude do we have,
\begin{align}
    (1-t/\lam_\infty)^{\alpha_q(s)}
    =
    (1-s/\lam_\infty)^{\alpha_q(t)}
    =
    q^{ \alpha_q(s) \alpha_q(t) }
\end{align}
As described in \autoref{sec:review}, this pre-factor is explicitly non-meromorphic and introduces branch cuts beginning at~${s,t = \lam_{\infty} = \frac{1}{1-q}}$. We recall, however, that we explicitly allowed for such non-meromorphic behavior in the pre-factor of our ansatz~\eqref{eq:gVen} so long as $\cW(s,t)$ had no zeros nor poles in the region ${|s|, |t| < \lam_\infty}$. For the $p$-type and $r$-type amplitudes, the crossing-symmetric guess~${\cW(s,t) = (1-t/\lam_\infty)^{\alpha(s)} (1-s/\lam_\infty)^{\alpha(t)}}$ adds further non-polynomial behavior to each residue which cannot be fixed by any other crossing symmetric factor. Hence, we conclude that we can only cancel the non-polynomial residues in the case of the Coon amplitude with ${q<1}$. We thus take,
\begin{align}
\label{eq:prefactor}
    \cW_{q<1}(s,t)
    &=
    q^{ \alpha_q(s) \alpha_q(t) }
    \qquad
    \text{and}
    \qquad
    \cW_{q \geq 1}(s,t)
    =
    \cW_p(s,t)
    =
    \cW_r(s,t)
    =
    1
\end{align}
since there is no way to construct a crossing-symmetric pre-factor which cancels the non-polynomial behavior of each residue for the $p$-type and $r$-type amplitudes.

\sm

Of all the spectra which solve the generalized Veneziano amplitude constraints~\eqref{eq:VenConstraints}, only the Coon spectra~\eqref{eq:qSoln} can be included in an infinite product amplitude with polynomial residues. Moreover, for ${q<1}$ polynomial residues can only be achieved by introducing the non-meromorphic pre-factor ${\cW_q(s,t)=q^{\alpha_q(s)\alpha_q(t)}}$.

\sm

The other generalized Veneziano amplitudes~$\cA_p(s,t)$ and~$\cA_r(s,t)$ do not have polynomial residues, but the non-polynomial behavior of their residues is captured by the universal factor ${(1-t/\lam_\infty)^{-N}}$. These residues can be expanded in~$t$ for all~${|t| < \lambda_\infty}$, resulting in infinite spin exchange on each massive pole as described in~\autoref{sec:intro}. By construction, the massless poles have finite spin exchange with~${\ell_{\text{max}} = 1}$ (from the residue~$1/t$ multiplied by the kinematic pre-factor $P_4 = \cO(t^2)$ described in~\autoref{sec:review}).

\subsection{Unitarity?}

Although the non-polynomial residues of~$\cA_p(s,t)$ and~$\cA_r(s,t)$ are novel, these amplitudes may still be interesting. Amplitudes with non-polynomial residues have appeared in the context of extremized EFT-hedron bounds~\cite{Caron-Huot:2020cmc, Arkani-Hamed:2020blm}. Moreover, it has been recently shown that amplitudes with non-polynomial residues may be unitary~\cite{Huang:2022mdb}. The unitarity properties of the Coon amplitudes were also recently studied in~\cite{Geiser:2022icl, Figueroa:2022onw, Chakravarty:2022vrp, Bhardwaj:2022lbz}. Here we shall begin a unitarity analysis of the generalized Veneziano amplitudes.

\sm

In a unitary theory, the residue of each pole of the four-point amplitude must have an expansion on the Gegenbauer polynomials with positive partial wave coefficients. This expansion is described in~\eqref{eq:partialwave}. Several useful properties of the Gegenbauer polynomials are listed in the appendix of~\cite{Geiser:2022icl}. One particularly useful property is that the product of two Gegenbauer polynomials has a positive expansion on the Gegenabauer polynomials.

\sm

The unitarity properties of a given theory may depend on the number of spacetime dimensions~$d$. The Coon amplitudes exhibit a particularly rich dimension-dependence~\cite{Figueroa:2022onw}. For~${q>1}$, the Coon amplitude is non-unitary in any dimension. For~${0 < q \leq \frac{2}{3}}$, the Coon amplitude is unitary in any dimension. For~${\frac{2}{3} < q \leq 1}$, the Coon amplitude is unitary below a $q$\nobreakdash-dependent critical dimension $d_c(q)$. At ${q=1}$, this critical dimension ${d_c(1)=10}$ reproduces the critical dimension of the superstring. We shall derive similar results for the larger space of generalized Veneziano amplitudes. 

\subsubsection{Analytic results}

We begin with a dimension-agnostic analysis which will provide sufficient but not strictly necessary conditions for unitarity. We define ${z = \cos \theta}$, where $\theta$ is the scattering angle in the center-of-mass frame, so that $t = \half s (z-1)$. In terms of $z$, the residue of the generalized Veneziano amplitude at the massive pole ${s = \lambda_N}$ is given by,
\begin{align}
    \Res_{s=\lam_N}
    \cA(s,t)
    &=
    \cW \big( \lam_N, \half \lam_N (z-1) \big) \,
    \frac{ 1 }{ \lam_N }
    \bigg( 1 - \frac{\lam_N}{\lam_\infty} \bigg)
    \bigg( 1 - \frac{\lam_N}{\lam_{-\infty}} \bigg)
\\
    & \quad
    \times
    \frac{1}{ \big( 1-\half (z-1) \lam_N / \lam_\infty \big)^N }
    \prod_{n=1}^{N-1}
    \bigg[
    \bigg(
    \frac{1}{\lambda_n}
    - \frac{1}{\lambda_{\infty}}
    - \frac{1}{\lambda_{-\infty}}
    \bigg)
    \frac{\lam_N}{2}
    (z-1)
    + 1
    \bigg]
\no
\end{align}
The $z$\nobreakdash-independent factor,
\begin{align}
    \frac{ 1 }{ \lam_N }
    \bigg( 1 - \frac{\lam_N}{\lam_\infty} \bigg)
    \bigg( 1 - \frac{\lam_N}{\lam_{-\infty}} \bigg)
\end{align}
is always a positive number. The $z$\nobreakdash-dependent factor,
\begin{align}
    \frac{ \cW \big( \lam_N, \half \lam_N (z-1) \big) }
         { \big( 1-\half (z-1) \lam_N / \lam_\infty \big)^N }
\end{align}
has a positive expansion on the Gegenbauer polynomials since positive powers of~$z$ have a positive expansion on the Gegenbauer polynomials~\cite{Geiser:2022icl}. In the case of the Coon amplitude (for any $q$), this factor simply equals one. In all other cases, the factor ${\cW(s,t) = 1}$, and we may use the binomial theorem to write,
\begin{align}
    \frac{ 1 }{ \big( 1-\half (z-1) \lam_N / \lam_\infty \big)^N }
    =
    \bigg(
    1 + \frac{\lam_N}{2 \lam_\infty}
    \bigg)^{-N}
    \sum_{k=0}^\infty
    \binom{N+k-1}{k}
    \bigg(
    \frac{ \lam_N }{ \lam_N + 2\lam_\infty }
    \bigg)^k
    z^k
\end{align}
which is a sum of powers of $z$ with manifestly positive coefficients. It remains to study the polynomial part of the residue,
\begin{align}
    \prod_{n=1}^{N-1}
    \bigg[
    \bigg(
    \frac{1}{\lambda_n}
    - \frac{1}{\lambda_{\infty}}
    - \frac{1}{\lambda_{-\infty}}
    \bigg)
    \frac{\lam_N}{2}
    \, z
    -
    \bigg(
    \frac{1}{\lambda_n}
    - \frac{1}{\lambda_{\infty}}
    - \frac{1}{\lambda_{-\infty}}
    \bigg)
    \,
    \frac{\lam_N}{2}
    + 1
    \bigg]
\end{align}
This factor will be a sum of powers of $z$ with manifestly positive coefficients if,
\begin{align}
    \frac{1}{\lambda_n}
    - \frac{1}{\lambda_{\infty}}
    - \frac{1}{\lambda_{-\infty}}
    \geq 0
    \qquad
    &\text{and}
    \qquad
    \bigg(
    \frac{1}{\lambda_n}
    - \frac{1}{\lambda_{\infty}}
    - \frac{1}{\lambda_{-\infty}}
    \bigg)
    \frac{\lam_N}{2}
    \leq 1
\intertext{for ${n=1,2,\dots,N-1}$ for each ${N \geq 1}$. Because the poles are ordered $\lam_{n} > \lam_{n-1}$, these conditions are satisfied for all $n$ at fixed $N$ if,}
    \frac{1}{\lambda_{N-1}}
    - \frac{1}{\lambda_{\infty}}
    - \frac{1}{\lambda_{-\infty}}
    \geq 0
    \qquad
    &\text{and}
    \qquad
    \bigg(
    1
    - \frac{1}{\lambda_{\infty}}
    - \frac{1}{\lambda_{-\infty}}
    \bigg)
    \frac{\lam_N}{2}
    \leq 1
\intertext{where we have used ${\lam_1 = 1}$. These conditions are in turn satisfied for all $N$ if,}
    \frac{1}{\lambda_{\infty}}
    - \frac{1}{\lambda_{\infty}}
    - \frac{1}{\lambda_{-\infty}}
    \geq 0
    \qquad
    &\text{and}
    \qquad
    \bigg(
    1
    - \frac{1}{\lambda_{\infty}}
    - \frac{1}{\lambda_{-\infty}}
    \bigg)
    \frac{\lam_\infty}{2}
    \leq 1
\intertext{Rearranging, we find,}
\label{eq:UniConds}
    \frac{1}{\lam_{-\infty}}
    \leq 0
    \qquad
    &\text{and}
    \qquad
    \frac{3}{\lam_{\infty}}
    + \frac{1}{\lam_{-\infty}}
    \geq 1
\end{align}
We have carefully written these conditions in terms of the reciprocals $1/\lam_\infty$ and $1/\lam_{-\infty}$ since $\lam_\infty$ or $\lam_{-\infty}$ may be infinite.

\sm

The conditions~\eqref{eq:UniConds} are satisfied as follows. For the Coon amplitudes with ${q < 1}$, the first condition is trivially satisfied since ${1/\lam_{-\infty} = 0}$, leaving only the second condition,
\begin{align}
    \text{Coon} \, (q < 1) :
    &&
    3(1-q) &\geq 1 
    &
    \implies
    &&
    q &\leq \tfrac{2}{3}
\intertext{For the Coon amplitudes with ${q \geq 1}$, we have ${1/\lam_{\infty} = 0}$, and the two conditions become,}
    \text{Coon} \, (q \geq 1) :
    &&
    1-q &\leq 0
    &
    \implies
    &&
    q &\geq 1
\no \\[1ex]
    &&
    1-q &\geq 1
    &
    \implies
    &&
    q &\leq 0
\intertext{which is never satisfied. For the $p$-type amplitudes, the two conditions become,}
    p\text{-type} :
    &&
    \frac{ 1-x/p}{ 1+x \phantom{/p} }
    &\leq 0
    &
    \implies
    &&
    x &\geq p
\no \\[1ex]
    &&
    3 \, \frac{ 1-xp}{ 1+x \phantom{p} }
    + \frac{ 1-x/p}{ 1+x \phantom{/p} }
    &\geq 1
    &
    \implies
    &&
    x &\leq \frac{3p}{1+p+3p^2}
\intertext{Finally, for the $r$-type amplitudes, the two conditions become,}
    r\text{-type} :
    &&
    \frac{ 1-x }{ 1+x }
    &\leq 0
    &
    \implies
    &&
    x &\geq 1
\no \\[1ex]
    &&
    4 \, \frac{ 1-x }{ 1+x }
    &\geq 1
    &
    \implies
    &&
    x &\leq \frac{3}{5}
\end{align}
which is never satisfied.

\sm

\begin{figure}

\centering

\begin{tikzpicture}

\begin{axis}
[
    width = 1\textwidth,
    height = 0.7\textwidth,
    xmin = 0, xmax = 1.5,
    ymin = 0, ymax = 1.5,
    xtick = {0,2/3,1},
    ytick = {1},
    extra y ticks = {0},
    extra y tick labels = \empty,
    xticklabels = {$0$, $\tfrac{2}{3}$, $1$},
    xlabel = {$x$},
    ylabel = {$y$},
    x label style = {at={(current axis.right of origin)}, anchor=west},
    y label style = {at={(current axis.north west)}, rotate=-90, anchor=south},
    axis lines = left,
    legend cell align = {left},
    legend pos = north west,
    legend entries =
    {
        \, string spectrum,
        \, Coon spectra,
        \, $r$-type spectra,
        \, unitary in $d \leq 4$,
        \, unitary in $d \leq 6$,
        \, unitary in $d \leq 10$,
        \, unitary in all $d$,
        \, non-unitary,
        \, unphysical spectra
    }
]

\path[name path = xtop] (0,1.5) -- (1.5,1.5);
\path[name path = xbottom] (0,0) -- (1.5,0);

\addlegendimage{only marks, mark size = 2pt, fill = green}
\addlegendimage{ultra thick, color = black}    
\addlegendimage{ultra thick, color = black, densely dashed}
\addlegendimage
{
    only marks,
    mark = square*,
    color = green!40!white,
    mark size = 4pt
}   

\addlegendimage
{
    only marks,
    mark = square*,
    color = green!80!black,
    mark size = 4pt
}    

\addlegendimage
{
    only marks,
    mark = square*,
    color = green!45!black,
    mark size = 4pt
}

\addlegendimage
{
    only marks,
    mark = square*,
    color = blue!60,
    mark size = 4pt
}

\addlegendimage
{
    only marks,
    mark = square*,
    color = yellow!40,
    mark size = 4pt
}

\addlegendimage
{
    only marks,
    mark = square*,
    color = red!30,
    mark size = 4pt
}

\addplot
[
    name path = c,
    domain = 0:1.5, 
]
{x*x};

\addplot
[
    domain = 0.666:1.5, 
    ultra thick,
	black,
    smooth,
    opacity = 0.75
]
{x*x};

\addplot
[
    domain = 0:0.667, 
    ultra thick,
	blue!60,
    smooth,
]
{x*x};

\addplot
[
    name path = r,
    domain = 0:1, 
    ultra thick,
	black,
    densely dashed,
    smooth,
    opacity = 0.75
]
{x*(1+x)/(3-x)};

\addplot
[
    only marks,
    mark size = 3pt,
    fill = green
]
coordinates {(1,1)};

\addplot
[
    name path = d4,
    smooth,
    green!40!white,
    on layer = axis background
]
file[skip first]{d4unitary.dat};

\addplot
[
    name path = d6,
    smooth,
    green!80!black,
    on layer = axis background
]
file[skip first]{d6unitary.dat};

\addplot
[
    name path = d10,
    smooth,
    green!45!black,
    on layer = axis background
]
file[skip first]{d10unitary.dat};

\addplot
[
    samples = 500,
    name path = fm1,
    domain = 0:0.667,
    on layer = axis background,
	blue!60,
    smooth,
]
{ x*x*(6+x-3*x*x-sqrt(9-6*x-11*x*x))/(9-3*x-5*x*x+3*x*x*x) };

\addplot
[
    samples = 200,
    name path = fm2,
    domain = 0.666:0.672,
    on layer = axis background,
	blue!60,
    smooth,
]
{ x*x*(6+x-3*x*x-sqrt(9-6*x-11*x*x))/(9-3*x-5*x*x+3*x*x*x) };

\addplot
[
    samples = 200,
    name path = fp,
    domain = 0.666:0.672,
    on layer = axis background,
	blue!60,
    smooth,
]
{ x*x*(6+x-3*x*x+sqrt(9-6*x-11*x*x))/(9-3*x-5*x*x+3*x*x*x) };

\addplot
[
    name path = f0,
    domain = 0:0.667, 
    on layer = axis background,
	blue!60,
    smooth,
]
{ x*x };

\addplot[yellow!40]
fill between
[
    of = r and c,
    soft clip = {domain = 0:1},
    on layer = axis background
];

\addplot[yellow!40]
fill between
[
    of = c and xbottom,
    soft clip = {domain = 0:1.2247},
    on layer = axis background
];

\addplot[yellow!40]
fill between
[
    of = xtop and xbottom,
    soft clip = {domain = 1.224:1.5},
    on layer = axis background
];
 
\addplot[green!40!white] fill between [of=r and d4, on layer = axis background];
\addplot[green!80!black] fill between [of=r and d6, on layer = axis background];
\addplot[green!45!black] fill between [of=r and d10, on layer = axis background];

\addplot[blue!60] fill between [of=fm1 and f0, on layer = axis background];
\addplot[blue!60] fill between [of=fm2 and fp, on layer = axis background];

\addplot[red!30]
fill between
[
    of = r and xtop,
    soft clip = {domain = 0:1},
    on layer = axis background
];

\addplot[red!30]
fill between
[
    of = c and xtop,
    soft clip = {domain = 0.99:1.2247},
    on layer = axis background
];

\end{axis}

\end{tikzpicture}

\caption{The two-parameter space of solutions to the generalized Veneziano amplitude constraints with unitary regions in various dimensions. The blue region is unitary in all dimensions and includes the Coon amplitudes with ${0 < q \leq \frac{2}{3}}$. The dark green region is unitary in ${d \leq 10}$ and includes the Veneziano amplitude at ${x=y=1}$. The middle green region is unitary in ${d \leq 6}$. The light green region is unitary in ${d \leq 4}$. The blue region was computed analytically, and the green regions were computed by numerically analyzing the first few partial wave coefficients $c_{n,j}$.}

\label{fig:Ven2}

\end{figure}
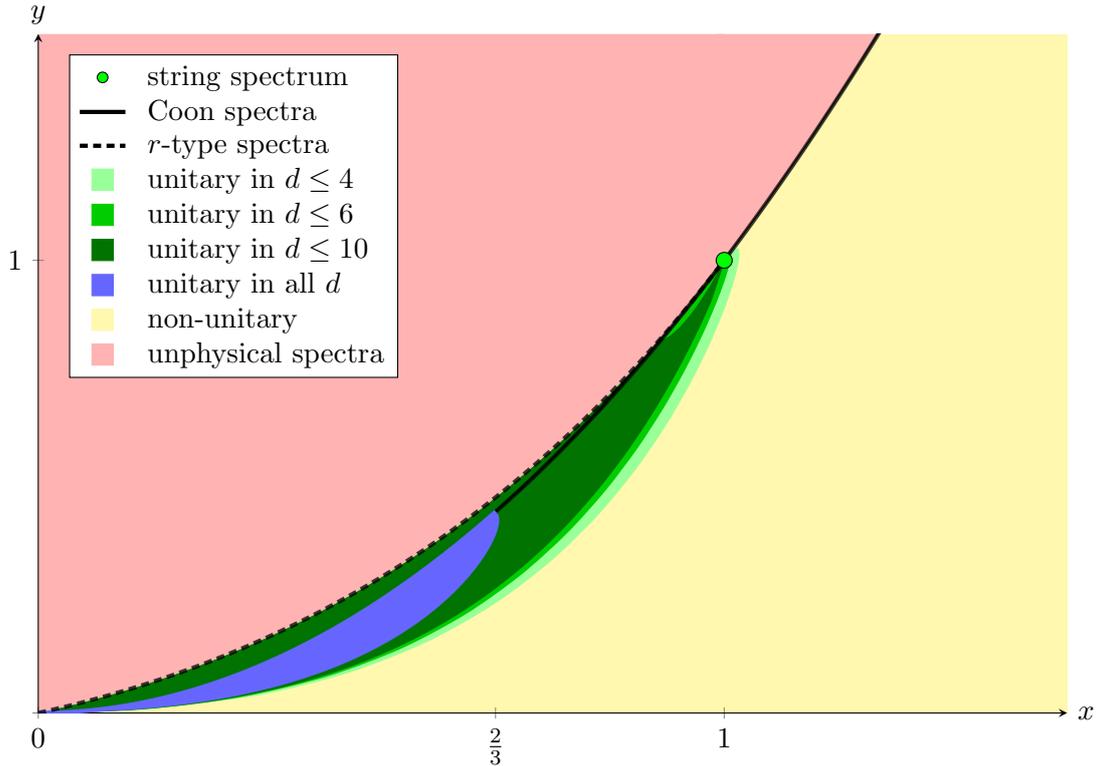

In summary, we have found that the Coon amplitudes with ${0 < q \leq \frac{2}{3}}$ are unitary in any dimension, in agreement with~\cite{Figueroa:2022onw}. Moreover, we have analytically demonstrated that the $p$-type generalized Veneziano amplitudes with ${p \leq x \leq 3p / (1+p+3p^2)}$ are unitary in any dimension. These inequalities define a region of parameter space with infinite critical dimension. In terms of the parameters $x$ and~$y$, the first inequality ${p \leq x}$ becomes~${y \leq x^2}$ while the second inequality becomes ${f_-(x) \leq y \leq f_+(x)}$, where,
\begin{align}
    f_{\pm}(x)
    =
    \frac{ x^2 ( 6+x-3x^2 \pm \sqrt{9-6x-11x^2} ) }
         { 9-3x-5x^2+3x^3 }
\end{align}
This infinite critical dimension region is displayed in~\autoref{fig:Ven2}.

\sm

While the conditions~\eqref{eq:UniConds} are sufficient to prove unitarity in all dimensions, they are by no means necessary. In general, for a given finite dimension $d$, the unitary region in the $xy$-plane will be larger than the region of infinite critical dimension.

\subsubsection{Numerical results}

To study the regions of parameter space with finite critical dimension, we shall employ numerical methods to analyze the first few partial wave coefficients. One cannot mathematically prove unitarity by examining a finite number of partial wave coefficients, but if any of those coefficients are negative, then the amplitude in question is non-unitary.

\sm

In this way, one can find evidence for the critical dimension of the superstring by computing one of the first partial wave coefficients of the Veneziano amplitude, ${c_{3,0} \propto 10-d}$. Similarly, one can deduce that the Coon amplitudes with~${q > 1}$ are non-unitary by computing the coefficient ${c_{2,0} \propto 1-q}$~\cite{Geiser:2022icl}. In this spirit, we hope to provide some evidence on the qualitative structure of the unitary regions of parameter space for the generalized Veneziano amplitudes.

\sm

The analytic expressions for the partial wave coefficients $c_{n,j}$ are given by the following overlap integral against the Gegenbauer polynomials~\cite{Geiser:2022icl},
\begin{align}
    c_{n,j}
    &=
    \cN^{(\frac{d-3}{2})}_j
    \int_{-1}^1 dz \,
    (1-z^2)^{\frac{d-4}{2}} \,
    C^{(\frac{d-3}{2})}_j(z)
    \times
    \Res_{s=\lam_n}
    \cA \big( s, \half s (z-1) \big)
\end{align}
where the residue is given by~\eqref{eq:Res} and the normalization is,
\begin{align}
    \cN^{(\frac{d-3}{2})}_j
    &=
    2^{d-5} (2j+d-3) 
    \frac{ \Gamma(j+1) \Gamma(\frac{d-3}{2})^2 }
         { \pi \Gamma(j+d-3) }
\end{align}
The apparent poles in this formula at~${d=3}$ are a remnant of the normalization of the Gegenbauer polynomials and can be trivially removed by a change in normalization.

\sm

For the $p$-type and $r$-type amplitudes which exhibit infinite spin exchange, the coefficients $c_{n,j}$ with ${n \geq 1}$ will generally be non-zero for all spins ${j \geq 0}$. Remarkably, Mathematica can explicitly compute these integrals in terms of generalized hypergeometric functions. The expressions are incredibly long, so we shall omit them here. Instead, we shall numerically examine the region of parameter space where ${c_{n,j} \geq 0}$ for ${1 \leq n \leq 4}$ and ${0 \leq j \leq 3}$ in several dimensions, namely $d=4,6,10$. The unitary regions of parameter space are displayed in~\autoref{fig:Ven2}. The qualitative structures of these regions do not appreciably change upon probing larger values of~$n$ or~$j$.

\sm

As expected, the unitary region in $d$ dimensions envelopes those in $d' > d$ dimensions, and they all contain the region of infinite critical dimension. For ${d \leq 10}$ the unitary region includes the Veneziano amplitude at ${x=y=1}$. The unitary region also appears to contain some $r$-type amplitudes, albeit with finite critical dimension. It would be interesting to study the features of~\autoref{fig:Ven2} in more detail. Perhaps the methods of~\cite{Geiser:2022icl, Figueroa:2022onw, Chakravarty:2022vrp, Bhardwaj:2022lbz} which were used to study the Coon amplitudes could be adapted to study the unitary properties of generalized Veneziano amplitudes.


\section{Generalized Virasoro amplitudes}
\label{sec:Vir}

In this section, we shall systematically analyze our infinite product ansatz~\eqref{eq:gVir} for the generalized Virasoro amplitude.

\subsection{Virasoro truncation}

We first recall the infinite product form~\eqref{eq:VirIP} of the Virasoro amplitude, which has simple poles at each non-negative integer. After applying the mass-shell relation ${s+t+u=0}$, the residue of the massless $s$\nobreakdash-channel pole is~${1/t^2}$, and the residue of each massive pole at~${s=N}$ is a polynomial of degree\nobreakdash-${(2N-2)}$ in~$t$. The Virasoro amplitude achieves these residues because on each $s$-pole, its zeros cancel the $t$\nobreakdash-poles and $u$\nobreakdash-poles, leaving a finite polynomial in~$t$. This cancellation can be described in terms of the numerator factors,
\begin{align}
    \cN_n(s,t,u)
    =
    1 + (st+tu+us)/n^2 + stu/n^3
\end{align}
When~$s=N$, each numerator factor ${\cN_{N+n}(N,t,-N-t) \propto (1-t/n)(1-u/n)}$ cancels both the $t$-channel and $u$-channel pole factors~${(1-t/n)^{-1} (1-u/n)^{-1}}$, and the infinite product truncates. In short, the condition,
\begin{align}
\label{eq:VirTrunc}
    \cN_{N+n}(N,n,-N-n) = 0
\end{align}
ensures that the Virasoro amplitude has polynomial residues.

\sm

These features bare a striking resemblance to those of the Veneziano amplitude. Hence, our analysis of the generalized Virasoro amplitude~\eqref{eq:gVir} will mirror our analysis of the generalized Veneziano amplitudes in the previous section.

\subsection{Generalized Virasoro truncation}

We now return to our generalized Virasoro ansatz~\eqref{eq:gVir}. We shall demand that the zeros and poles of this amplitude cancel in a similar fashion as those of the Virasoro amplitude. We first demand that the residue at~${s=0}$ is~${1/t^2}$ so that the amplitude reproduces the massless spectrum of supergravity analogously to the Virasoro amplitude,
\begin{align}
    \Res_{s=0}
    \cA(s,t,u)
    =
    \frac{1}{t^2}
    \qquad
    \implies
    \qquad
    \cW(0,t,-t) \,
    \prod_{n \geq 1}
    \frac{ 1 - \mask{t^2/\lam_n^2}{A_n t^2} }
         { 1 - t^2/\lam_n^2 }
    = 1
\end{align}
which implies that $\cW(0,t,-t) = 1$ and~$A_n = 1/\lam_n^2$ since~$\cW(s,t,u)$ has neither zeros nor poles. In other words, the coefficients~$A_n$ are again determined by the poles~$\lambda_n$.

\sm

Next, in analogy with the truncation condition for the Virasoro amplitude~\eqref{eq:VirTrunc}, we demand that the generalized numerator factor,
\begin{align}
    \cN_n(s,t,u)
    =
    1+A_n(st+tu+us)-B_nstu
\end{align}
obeys the generalized truncation condition,
\begin{align}
\label{eq:gVirTrunc}
    \cN_{N+n}(\lambda_N,\lambda_n,-\lambda_N-\lambda_n) = 0
\end{align}
so that ${\cN_{N+n}(\lam_N,t,-\lambda_N-t) \propto (1-t/\lam_n)(1-u/\lam_n)}$ and the infinite sequence of $t$-poles and $u$-poles cancels on each $s$-channel pole. This truncation condition determines the coefficients~$B_n$ in terms of the poles~$\lam_n$,
\begin{align}
    B_n
    &=
    \frac{  \lam_k^2
          + \lam_k^{\phantom{2}} \lam_{n-k}^{\phantom{2}}
          + \lam_{n-k}^2
          - \lam_{n \vphantom{k}}^2 }
         {  \lam_{n \vphantom{k}}^2 \lam_{n-k}^{\phantom{2}} \lam_k^{\phantom{2}}
          ( \lam_k^{\phantom{2}} + \lam_{n-k}^{\phantom{2}} ) }
    &
    k
    &=
    1,2,\dots,n-1
\end{align}
For fixed ${n \geq 2}$, both~$k$ and~${k'=n-k}$ yield the same equation for~$B_n$ so that there are again~$\floor{\frac{n}{2}}$ independent equations for~$B_n$. Once more, the coefficient~$B_1$ is left undetermined, the coefficients~$B_2$ and~$B_3$ are uniquely determined, and the coefficients~$B_n$ with~${n\geq 4}$ are all over-determined.

\sm

As in the previous section, this over-determination of the~$B_n$ highly constrains the poles. Any sequence of poles~$\lam_n$ must leave the following combination independent of~$k$ for all~${n \geq 2}$,
\begin{align}
\label{eq:VirConstraints}
    \Lambda_n(k)
    &=
    \frac{ \lam_k^2
         + \lam_k^{\phantom{2}} \lam_{n-k}^{\phantom{2}}
         + \lam_{n-k}^2
         - \lam_{n\vphantom{k}}^2 }
         { \lam_{n-k} \lam_k ( \lam_k+\lam_{n-k} ) }
\end{align}
We shall refer to these equations as the generalized Virasoro amplitude constraints. The Virasoro solution~${\lam_n=n}$ (i.e.\ the string theory spectrum) solves these constraints with ${\Lambda_n(k)=-1/n}$ for all~$n$ and~$k$. We shall now search for other, more general sequences of poles~$\lam_n$ which solve the generalized Virasoro amplitude constraints.

\subsection{Generalized Virasoro amplitude constraints}

Since~$\Lambda_n(k)$ must be independent of~$k$, we may fix~$n \geq 2$ and choose two distinct values of~$(k,\ell)$ in the appropriate range to find,
\begin{equation}
    \frac{ \lam_k^2
         + \lam_k^{\phantom{2}} \lam_{n-k}^{\phantom{2}}
         + \lam_{n-k}^2
         - \lam_{n\vphantom{k}}^2 }
         { \lam_{n-k} \lam_k (\lam_k + \lam_{n-k} ) }
    =
    \frac{ \lam_\ell^2
         + \lam_\ell^{\phantom{2}} \lam_{n-\ell}^{\phantom{2}}
         + \lam_{n-\ell}^2
         - \lam_{n\vphantom{\ell}}^2 }
         { \lam_{n-\ell} \lam_\ell (\lam_\ell + \lam_{n-\ell} ) }
\end{equation}
We may then solve this equation for~$\lam_n$ in terms of~$\lam_k$, $\lam_{n-k}$, $\lam_{\ell}$, and~$\lam_{n-\ell}$,
\begin{equation}
\label{eq:VirRec}
    \lam_n
    =
    \sqrt{
    \tfrac{ \lam_k^{\phantom{2}} \lam_{n-k}^{\phantom{2}}
          ( \lam_{n-k}^{\phantom{2}}
          + \lam_k^{\phantom{2}} )
          ( \lam_{n-\ell}^2
          + \lam_{n-\ell}^{\phantom{2}} \lam_{\ell}^{\phantom{2}}
          + \lam_{\ell}^2 )
          - \lam_{\ell}^{\phantom{2}} \lam_{n-\ell}^{\phantom{2}}
          ( \lam_{n-\ell}^{\phantom{2}}
          + \lam_{\ell}^{\phantom{2}} )
          ( \lam_{n-k}^2
          + \lam_{n-k}^{\phantom{2}} \lam_k^{\phantom{2}}
          + \lam_k^2) }
          { \lam_k \lam_{n-k}
          ( \lam_{n-k} + \lam_k )
          - \lam_{\ell} \lam_{n-\ell}
          ( \lam_{n-\ell} + \lam_{\ell} ) }
    }
\end{equation}
As in the generalized Veneziano case, the first three poles are free parameters, and~\eqref{eq:VirRec} determines all the subsequent poles in terms of~$\lambda_1$, $\lambda_2$, and~$\lambda_3$. We shall again define the positive numbers~${x=\lam_2-\lam_1=\lam_2-1>0}$ and~${y=\lam_3-\lam_2>0}$, using the fact that the poles~${\lam_n > \lam_{n-1}}$ are ordered. The choice~${\lam_1 = 1}$ simply sets our units.

\sm

For~${n=4}$ and~${n=5}$, there is a unique choice of $(k,\ell)$ and thus a single equation determining $\lambda_4 = \lambda_4(x,y)$ and $\lambda_5 = \lambda_5(x,y)$. For~${n=6}$, we can write two different equations for~${\lambda_6 = \lambda_6(x,y)}$. These equations are exceedingly large and include several nested radicals. Equating these two expressions implicitly defines a curve in the $xy$-plane. Any solution of the generalized Virasoro amplitude constraints must be on this curve. We have analyzed this curve numerically and verified that it passes through~${x=y=1}$. Repeating this process at~${n=7}$ yields a second curve in the $xy$-plane, and any solution of the generalized Virasoro amplitude constraints must again be on this curve. Through a straightforward numerical analysis, we find that the~$\lambda_6$ and~$\lambda_7$ curves only intersect at~${x=y=1}$, corresponding to the string spectrum.

\sm

In other words, only the string spectrum~${\lam_n=n}$ satisfies the generalized Virasoro amplitude constraints~\eqref{eq:VirConstraints}. Thus, the construction which led to several infinite families of generalized Veneziano amplitudes fails to yield any new generalizations of the Virasoro amplitude. The closed string is highly constrained.


\section{Discussion}
\label{sec:disc}

In this paper, we have systematically analyzed generalizations of both the Veneziano and Virasoro amplitudes by considering the infinite product ansatz~\eqref{eq:gVen} and~\eqref{eq:gVir}. Demanding that the poles cancel on each residue, we arrived at the generalized Veneziano and generalized Virasoro amplitude constraints,~\eqref{eq:VenConstraints} and \eqref{eq:VirConstraints}, respectively. These constraints are equivalent to an infinite set of non-linear recursion relations obeyed by the poles of each amplitude.

\sm

In the generalized Veneziano case, we solved the recursion relations analytically by reducing them to the Riccati recursion relation~\eqref{eq:Riccati}. The solutions corresponded to the Veneziano amplitude, the one-parameter family of Coon amplitudes, and a larger two-parameter family of amplitudes with an infinite tower of spins at each mass level. Of these generalized Veneziano amplitudes, only the Veneziano and Coon amplitudes have polynomial residues. We also began an initial study of the unitarity properties of these amplitudes and found that a subspace of them, including the Coon amplitudes with~${0 < q \leq \frac{2}{3}}$, are unitary in any dimension. A larger subspace is unitary with finite critical dimension.

\sm

In the generalized Virasoro case, we numerically demonstrated that the only consistent solution to the generalized Virasoro amplitude constraints is the string spectrum. These infinitely many constraints did not allow any deviation outside of closed string theory. Our results are consistent with those of~\cite{Geiser:2022icl, Cheung:2022mkw}.

\sm

In future work, it would be interesting to explore where the low-energy expansion coefficients of the generalized Veneziano amplitudes~$\cA_p(s,t)$ and~$\cA_r(s,t)$ lie in relation to the EFT-hedron~\cite{Arkani-Hamed:2020blm} and other positivity bounds~\cite{Caron-Huot:2020cmc}. The low-energy expansion coefficients of the Coon amplitudes were recently studied in this context in~\cite{Figueroa:2022onw, Geiser:2022icl}. It would also be interesting to further study the unitarity properties of these amplitudes. Further generalizations of this work may study other truncation conditions leading to polynomial residues for our infinite product ansatz. Recent progress in this direction has been made in~\cite{Cheung:2022mkw}.

\sm

Finally, we hope to find a definitive field theory or string theory realization of the Coon amplitudes or their generalizations. Recently, accumulation point spectra like those exhibited by Coon amplitudes were found in a setup involving open strings ending on a D-brane~\cite{Maldacena:2022ckr}. Moreover, accumulation point spectra have appeared in various contexts in the modern S-matrix bootstrap program, so it is imperative to better understand the Coon amplitudes' physical origins.


\bibliography{GenVAmps}

\end{document}